\documentclass[12pt,preprint]{aastex}

\usepackage{lscape}
\usepackage{natbib}

\begin{document}

\title{Identifying Contributions to the Stellar Halo from Accreted, Kicked-Out, and In Situ Populations}

\author{Allyson A. Sheffield\altaffilmark{1}, Steven R. Majewski\altaffilmark{2}, 
Kathryn V. Johnston\altaffilmark{1}, Katia Cunha\altaffilmark{3,4,5}, 
Verne V. Smith\altaffilmark{4}, Andrew M. Cheung\altaffilmark{2,6}, Christina M. Hampton\altaffilmark{2}, 
Trevor J. David\altaffilmark{7}, Rachel Wagner-Kaiser\altaffilmark{8}, Marshall C. Johnson\altaffilmark{9}, Evan Kaplan\altaffilmark{10}, Jacob Miller\altaffilmark{10}, and Richard J. Patterson\altaffilmark{2}}

\altaffiltext{1}{Department of Astronomy, Columbia University, Mail Code 5246, 
New York, NY 10027 (asheffield, kvj@astro.columbia.edu)}

\altaffiltext{2}{Department of Astronomy, University of Virginia, P.O. Box 400325, 
Charlottesville, VA 22904 (srm4n, rjp0i@virginia.edu)}

\altaffiltext{3}{Observatorio Nacional, Rio de Janeiro, Brazil}

\altaffiltext{4}{National Optical Astronomy Observatories, PO Box
26732, Tucson, AZ 85726 (cunha, vsmith@noao.edu)}

\altaffiltext{5}{Steward Observatory, University of Arizona, Tucson, AZ 85721, USA}

\altaffiltext{6}{Department of Materials Science and Engineering, University of Virginia, P.O. Box 400745, 
Charlottesville, VA 22904 (amc4e@virginia.edu)}

\altaffiltext{7}{Department of Astronomy, California Institute of
Technology, 1200 E. California Blvd., MC 249-17, Pasadena, CA 91125
(tjd@astro.caltech.edu)}

\altaffiltext{8}{Department of Astronomy, University of Florida, 211 Bryant Space Science Center, Gainesville, FL 32611 (rawagnerkaiser@astro.ufl.edu)}

\altaffiltext{9}{Department of Astronomy, The University of Texas at Austin, 1 University Station, C1400, Austin, TX 78712 (mjohnson@astro.as.utexas.edu)} 

\altaffiltext{10}{Department of Physics and Astronomy, Vassar College, 124 Raymond Ave., Poughkeepsie, NY 12604 (evakaplan@vassar.edu)}

\begin{abstract}
We present a medium-resolution spectroscopic survey of late-type giant stars at mid-Galactic latitudes of (30$^{\circ}<|b|<$60$^{\circ}$), designed to probe the properties of this population to distances of $\sim$9 kpc.
Because M giants are generally metal-rich and we have limited contamination from thin disk stars by the latitude selection, most of the stars in the survey are expected to be members of the thick disk ($<$[Fe/H]$>\sim$-0.6) with some contribution from the metal-rich component of the nearby halo.  

Here we report first results for 1799 stars.
The distribution of radial velocity (RV) as a function of $l$ for these stars shows (1) the expected thick disk population
and (2) local metal-rich halo stars moving at high speeds relative to the disk, that in some cases form distinct sequences in RV-$l$ space. 
High-resolution echelle spectra taken for 34 of these ``RV outliers'' reveal the following patterns across the [Ti/Fe]-[Fe/H] plane: seventeen of the stars have 
abundances reminiscent of the populations present in dwarf satellites of the Milky Way;
eight have abundances coincident with those of the Galactic disk and more metal-rich halo;
and 
nine of the stars fall on the locus defined by the majority of stars in the halo.
The chemical abundance trends of the RV outliers suggest that this sample consists predominantly of stars accreted from infalling dwarf galaxies. 
A smaller fraction of stars in the RV outlier sample may have been formed in the inner Galaxy and subsequently kicked to higher eccentricity orbits, but the sample is not large enough to distinguish conclusively between this interpretation and the alternative that these stars represent the tail of the velocity distribution of the thick disk. 
Our data do not rule out the possibility that a minority of the sample could have formed from gas {\it in situ} on their current orbits.
These results are consistent with scenarios where the stellar halo, at least as probed by M giants, arises from multiple formation mechanisms;
however, when taken at face value, our results for metal-rich halo giants suggest a much higher proportion to be accreted than found by \citet{carollo07,carollo10} and more like the fraction suggested in the analysis by \citet{nissen10,nissen11} and \citet{schuster12}.
We conclude that M giants with large RVs can provide particularly fruitful samples to mine for accreted structures and
that some of the velocity sequences may indeed correspond to real physical associations resulting from recent accretion events.  
\end{abstract}


\section{Introduction}
\subsection{Motivation for a Survey of Bright M Giants}
The formation history of the Milky Way is recorded in the present motions and chemical abundances of its stars.
Ideally, to unravel the Milky Way's history, we would like a catalog containing spatial, kinematical, and chemical data for every star in the Galaxy. 
Large-scale photometric surveys, such as the Two Micron All Sky Survey (2MASS) and the Sloan Digital Sky Survey (SDSS), are bringing us closer to this goal: both have led to sweeping views of the structure of the Galaxy.
Star counts from these surveys allow for detailed studies of the structure of each Galactic component \citep[e.g.,][]{skrutskie,maj03,bell08,juric,ivezic08} and are rich sources for follow-up spectroscopic studies \citep[e.g.,][]{maj04b,yanny09}.

The present study looks at the spectroscopic properties of a sample of relatively nearby ($d\lesssim$ 9 kpc) M giant stars at mid-Galactic latitudes of $30^{\circ}<|b|<60^{\circ}$ selected from the 2MASS catalog. 
M giants (1) are intrinsically bright stars and, hence, can be easily observed to large distances using even small telescopes; (2) can be readily identified on the basis of near-infrared, $JHK$ photometry and complete samples can be culled from full-sky catalogs (e.g., 2MASS); and (3) are limited to more metal-rich populations. These unique properties of M giants, when combined with the adopted latitude and magnitude selection criteria, make our survey particularly useful for exploring the structure of the thick disk and the metal-rich component of the nearby halo. 

In this first paper describing our survey, we present the photometric and spectroscopic data for our current thick-disk dominated sample of 1799 stars, but focus on the detection and interpretation of those stars that do not kinematically conform to the typical behavior of the Milky Way thick disk population and are likely to be members of the stellar halo. 
This stellar halo sample is unique compared to other halo surveys of stars in that it covers an intermediate distance range ($d <$ 9 kpc) and concentrates on a relatively rare halo tracer. 
For example, our survey volume is wider than that of the Hipparcos survey of the Sun's closest neighbors \citep[$d_{\rm lim,Hip}\sim$100 pc -][]{perryman97} as well as kinematically-selected solar neighborhood surveys \citep{els,carney96,schuster12} but is a much more local view of the stellar halo than the SDSS F-G type turnoff stars \citep[visible to 20-40 kpc -][]{juric,bell08} or the entire 2MASS M giant catalog \citep[visible out to 100 kpc -][]{maj03}. 
Our catalog is comparable in probed distances to the spectroscopic studies of the Century Survey Galactic Halo Project \citep{brown08}, RAVE \citep{steinmetz}, and \citet{carollo07,carollo10}, although these studies employed different, more commonly used halo tracers: blue horizontal branch (BHB) stars; stars -- both dwarfs and giants -- with $9<I<12$; and main sequence turnoff (MSTO) stars or dwarfs from the SDSS DR7 calibration sample, respectively.

\subsection{An M Giant Survey of the Nearby Galactic Halo}
In this first analysis of our survey we explore the origin of the M giant population in the nearby stellar halo. There are three broad categories of formation scenarios typically postulated for halo stars:
\begin{enumerate}
\item{\it In-situ-halo} stars form at comparable radii to their current locations within the dominant dark matter halo progenitor of the Galaxy. For example \citet{els} envisioned the stellar halo could be formed from infalling gas, prior to the formation of the disk, during an early monolithic collapse phase for our Galaxy, as seen in the hydrodynamical simulations of \citet{samland03}.
\item{\it Kicked-out} stars are formed initially more concentrated towards the center of the dominant dark matter halo,
within either the bulge or disk,
and are subsequently ejected to more eccentric orbits through minor or major mergers, as proposed by \citet{purcell10} and seen in hydrodynamical simulations by \citet{zolotov09} and \citet{mccarthy12}.\footnote{There are also two classes of stars, hypervelocity and runaway O-B stars, that are similar to this populaton \citep[see][and references within]{brown06}.  However, these are due to rare events within a binary system (e.g., ejection due to collisions within the system or an interaction with the Milky Way's central supermassive black hole) and we would not expect these classes to contribute a large fraction of halo stars.}
\citep[Note that there were many earlier studies of this same process that focused on the thickening or destruction of disks rather than the formation of the stellar halo, e.g.,][]{quinn93,walker96} 
\item{\it Accreted} stars form in separate dark matter halos that later merge with the Galaxy \citep[as proposed by e.g.,][]{sz78}. 
\end{enumerate}
Note that in prior theoretical work, the first two categories were both simply termed {\it in situ} to indicate more generally stars formed in the dominant galaxy dark matter halo progenitor \citep[as in, e.g.,][]{abadi06}. Separate terms are introduced here for clarity.

Many prior studies have probed these formation mechanisms by looking at the distribution of stars in different dimensions of phase-space.
For example, the importance of an {\it accreted} population has traditionally been assessed by looking for residual groupings that are signatures of stars' original associations whereas {\it in-situ-halo} or {\it kicked-out} populations are expected to be more smoothly distributed.
All-sky photometric surveys have revealed rich substructure in the outer halo (e.g., Sgr tidal tails - Majewski et al. 2003; the Anticenter Tributaries - Grillmair 2006; the Orphan stream - Belokurov et al. 2007) that can best be explained by accretion events \citep[e.g.,][]{bullock01,bullock05,bell08,johnston08,cooper10}.
In contrast, merger debris in the {\it nearby} halo fully phase-mix on a much shorter timescale, leading to the expectation of negligible evidence for accretion identifiable as coherent spatial structures \citep[e.g.,][]{johnston08,sharma10} and requiring additional dimensions of phase-space data to distinguish formation scenarios in this region.
Adding the dimension of line-of-sight velocities helps: conservation of phase-space density during phase-mixing requires that as debris from accretion events becomes less dense with time, it should become colder in velocity \citep[Liouville's Theorem; see e.g.,][]{helmi99a} and coherent structures may be apparent even once stars are smoothly distributed in space.
Indeed, substructure in velocities is apparent statistically over a large volume in BHB stars in the SDSS survey  (and shown to be broadly consistent with model stellar halos built within a hierarchical cosmology  --- see Xue et al. 2010, Cooper et al. 2011) and individual clumps in velocity have also been detected in the halo using metal-poor MSTO stars in the SEGUE survey \citep{schlaufman11}, K giant stars \citep{maj04a}, as well as in the mixed populations in RAVE \citep{williams}. 
Overall, group finding is more effective if even more dimensions of phase-space can be measured.
For example, \citet{maj92} analyzed proper motions for a sample of 250 F-K dwarfs in the direction of the North Galactic Pole that probes out to roughly 8 kpc; he measured a mean retrograde rotational velocity for the halo sample and detected a more coherent retrograde group of stars at a mean height above the Galactic plane of $Z\sim4.6$ kpc -- findings suggestive of an accreted halo population.  
Subsequently, \citet{mmh94,mmh96} obtained radial velocities for a subsample of stars from the \citet{maj92} survey and found a significant amount of phase-space clumpiness in their halo sample.
Once all six phase-space dimensions are known, conserved (or nearly conserved) quantities (such as energy, angular momentum, or orbital frequencies) can be calculated that can link stars from a common progenitor in the volume even if they are not at the same orbital phase, as has been done successfully for several nearby surveys \citep[see][]{helmi99a,helmi00,morrison09,gomez10}.

While the findings in the previous paragraph point to a significant fraction, and possibly the majority of the stellar halo being {\it accreted}, some contributions from {\it in-situ-halo} and {\it kicked-out} stars are still possible.
Structural, orbital and/or metallicity trends in the stellar halo with radius --- as well as transitions in those trends --- have historically been taken as indicative of these different formation mechanisms and leading to ``dual halo'' models \citep[e.g., in RR Lyraes, globular clusters and BHB stars, see][respectively]{hartwick87,zinn93,zinn96,sommerlarsen97}.
\citet{chiba2000} analyzed a kinematically unbiased sample of 1203 stars with [Fe/H]$<$-0.6 in the inner halo (within 4 kpc of the Sun) and detected a 
 gradient in the rotational velocity
 as a function of height above the Galactic plane for the more metal-rich stars in their sample 
--- as seen for the {\it in-situ-halo} stars formed in the simulations of \citet{samland03}.
\citet{chiba2000} also confirm the existence of the streams detected by \citet{helmi99b} --- further supporting some presence of an {\it accreted} population in this region.
\citet{carollo07,carollo10} studied the kinematics and metallicities of SDSS calibration stars for a 
larger volume (20 kpc)
and found evidence for two populations: one of metal-rich stars on only mildly eccentric orbits (which they dubbed ``inner halo'') and a second of metal-poor stars on more eccentric orbits (which they dubbed ``outer halo'').
Similarly, \citet{deason11} analyzed BHB stars in SDSS (a sample that probes the outer halo to 40 kpc) and find a net retrograde rotation for metal-poor stars and a net prograde rotation for metal-rich stars.
Such multi-component halos with distinct formation mechanisms for each component 
emerge naturally in the hydrodynamical simulations, with 
the inner halo (within 10-15 kpc) coming predominantly from 
stars (either {\it in-situ-halo} or {\it kicked-out}) formed within the main Galaxy dark matter halo progenitor and
the outer halo (dominant beyond 15-20 kpc) from mergers and accretion of stars by the Galactic dark matter halo \citep{abadi06,zolotov09,font11,mccarthy12,tissera12}.  
However, whether the distinct components observed in the stellar halo are indeed due to distinct formation mechanisms, and not merely a variety of accretion events, has yet to be proven.

In our own survey, the sample of stars is sufficiently far that distance and proper motion errors from extant data are too large to estimate their orbital properties accurately.
However, chemical abundances can be derived from high-resolution spectra in general, and thus provide an alternate avenue for exploring the origins of the stars.  
A star is branded at its birth by its chemical abundance patterns --- a signature that is generally conserved throughout its lifetime (like orbital properties) and cannot be diluted by orbital phase-mixing.
Moreover, stars deriving from a common origin should have similarities in their chemical abundance patterns, trends that are directly correlated to the details of their enrichment history.
Hence we can hope to ``chemically tag'' stars as members of the different populations via their abundance patterns (similar in spirit to Freeman \& Bland-Hawthorn's 2002 proposal for the reconstruction of ancient star-clusters in the stellar disk). 
The potential power of chemical tagging has already been demonstrated empirically in observations: in dwarf galaxies, for example, stars tend to have lower [$\alpha$/Fe] at a given [Fe/H] than stars in the bulk of the Milky Way's stellar halo \citep{smecker02,shetrone03,tolstoy04,geisler05,monaco07,chou10a} and similar patterns have been seen in
stars in stellar structures such as the Monoceros Ring \citep{chou10b} and Triangulum-Andromeda Cloud \citep{chou11}, which lends support to the interpretation of such features as originating from disrupted dwarf galaxies.
In a similar manner, a series of papers \citep{nissen10,nissen11,schuster12} have separated a sample of 94 nearby (within $\sim$335 pc of the Sun), metal rich (-1.6 $<$ [Fe/H] $<$ -0.4) stars into ``$\alpha$-rich'' and ``$\alpha$-poor'' groups and shown systematic differences in the abundances, ages and orbital properties of stars in these two groups that are suggestive of {\it kicked-out} and {\it accreted} origins, respectively.

\begin{figure}
\plotone{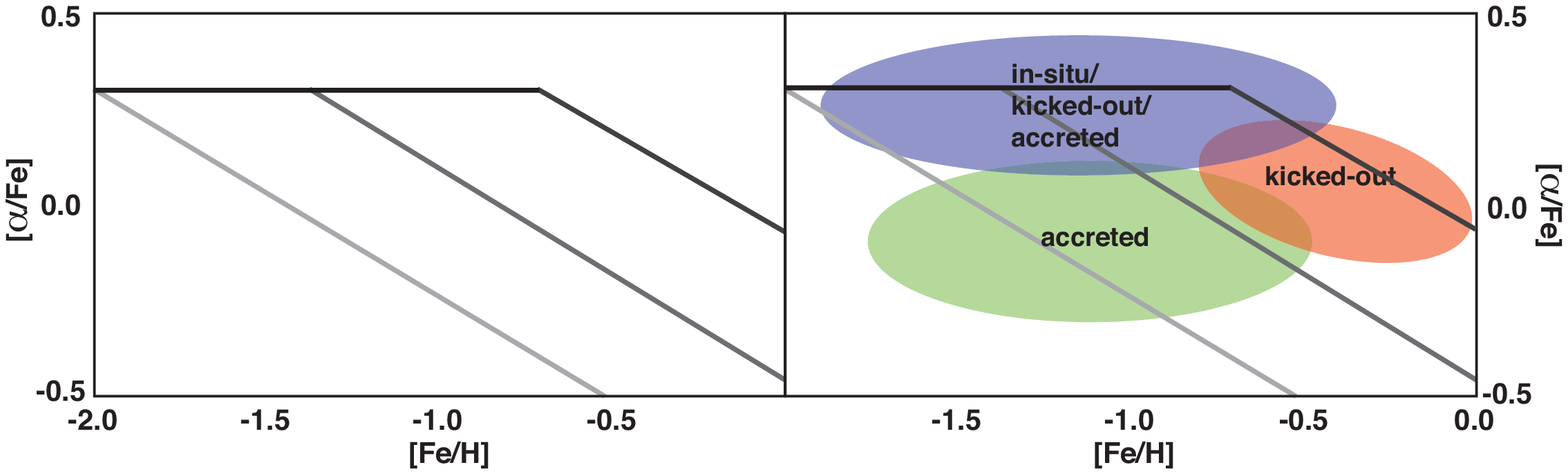}
\caption{\label{cartoon}Schematic showing the location in the [$\alpha$/Fe] vs. [Fe/H] plane for populations of stars originating from gas reservoirs with differing star formation efficiencies.  The left panel shows lines for three low/intermediate/high mass systems (assumed to represent stellar populations in small/medium/large dark matter halos), indicated by the light/medium/dark gray lines. All three systems are taken to have the same initial [$\alpha$/Fe] value due to Type II SNe enrichment; however, the point at which [$\alpha$/Fe] begins to decrease after the onset of Type Ia SNe explosions varies, with systems having the lowest (e.g., dwarf satellites) declining in [$\alpha$/Fe] at the lowest [Fe/H].  The right panel shows schematically the expected locations for three populations of halo stars within the Galaxy: {\it in-situ}, {\it kicked-out}, and {\it accreted}.}
\end{figure}

Figure \ref{cartoon} shows two cartoons to illustrate conceptually how this approach might be applied to our own sample.
The lines in the left hand panel show the expected temporal evolution, in the [$\alpha$/Fe]-[Fe/H] plane, of the gaseous chemical abundance for systems with low/intermediate/high star formation efficiencies (SFEs - indicated by lines with increasingly dark shades of gray), which are assumed to correspond to stellar systems embedded in small/intermediate/large dark matter potential wells \citep[see][]{mcwilliam97,gilmore98,robertson05}.
These systems could represent, for example: a Milky Way dwarf spheroidal (low SFE corresponding to low mass accreted systems), the LMC (intermediate SFE corresponding to intermediate mass accreted systems), and a Milky Way progenitor (high SFE and contributing to populations either formed {\it in situ} or {\it kicked out} from their original birth places).
All of these systems are expected to have old stars with high [$\alpha$/Fe] at low metallicity, which reflect the yields from explosive Type II SNe alone.
 Stars formed after the (delayed) onset of Type Ia SNe will become progressively more enriched by Fe and acquire lower [$\alpha$/Fe] as Type Ia SNe produce $\alpha$-elements much less efficiently.
The transition point in Fe between the early and late stages reflects the metallicity that the gas has reached prior to the onset of Type Ia enrichment, which will be lower/higher for systems that have less/more rapidly converted their gas into stars. 
 
The right hand panel of Figure \ref{cartoon} applies this intuition to show where populations with different origins might fall in this plane (these regions are defined more rigorously and empirically in Figure \ref{tag} and discussed in \S\ref{interp} using data from previous studies).
The region outlined in blue should contain stars formed early (because of their low metallicities and high [$\alpha$/Fe]) but that can now be in the halo via any of the three mechanisms ({\it in-situ-halo}, {\it kicked-out}, or {\it accreted}), and thus we do not anticipate being able to conclusively deduce their origins from this particular type of analysis.
The region outlined in green is likely only to contain stars formed more recently (because of the low [$\alpha$/Fe]) in small potential wells (because of the low Fe) and hence should be sensitive to a purely {\it accreted} population.
The region outlined in orange is likely only to contain stars formed more recently (because of the low [$\alpha$/Fe]) in deep potential wells (because of the high Fe) and hence should be {\it in}sensitive to {\it in-situ-halo} populations (formed exclusively early on) as well as the majority of {\it accreted} populations (because they form in smaller potential wells --- although, note, there could be some contamination from high-mass, late accreted systems, as suggested in Figure \ref{tag}). 
Stars that have formed recently in the Galactic disk and that were subsequently {\it kicked-out} might lie in this region
\cite[see][for an illustration of this idea with hydrodynamical cosmological simulations]{zolotov10}.
Overall, these expectations lead us to conclude that the fraction of our halo stars in the metal-poor and  $\alpha$-poor region is an indicator of the importance of late accretion, while the fraction with 
 disk-like abundances but moving at large speed relative to the Sun is indicative of the contribution of a recently {\it kicked-out} population.

Thus an additional advantage of focusing on abundance space as a probe of the relative proportions by origin of M giants in the stellar halo is that we only need to look for the expected systematic differences in the chemical composition of these populations ({\it in-situ-halo}, {\it kicked-out}, and {\it accreted}) overall rather than (for example) search for kinematical groupings from individual accretion events.
Hence, we can explore formation scenarios with much smaller samples of stars than by looking at dynamics alone.
Motivated by the promise of chemical abundances, 
which we could combine with the known RVs we already have for all program stars, we obtained high-resolution echelle spectra for 34 stars selected for follow-up based upon their high, halo-like speeds relative to the disk.
As a control sample we also observed five random thick disk red giants (chosen based on RVs similar to the bulk thick disk trend) and four M giant calibration stars from \citet{sl85} and \citet{smith00}.

This paper is organized in the following way.  
In \S\ref{prog}, we describe target selection, observations, and data reduction for the medium-resolution spectroscopy program of bright M giants.
The RVs and the identification 
of RV outliers -- stars that have high speeds relative to the bulk thick disk trend -- are presented in \S\ref{rvs}.
Details of the high-resolution follow-up spectroscopy program are given in \S\ref{hires}.
Our interpretation of the abundance data in terms of formation scenarios for the halo is given in \S\ref{interp}.
Lastly, we give a summary of the results and discuss them in the context of prior studies in \S\ref{conc}.
In a companion paper \citep{johnston12}, we develop a more complete understanding of the nature and implications of some possible dynamical groups in our survey by generating and analyzing synthetic observations of simulated stellar halos. 

\section{\label{prog}Program Stars}
\subsection{Defining the Sample}
M giants can be distinguished from M dwarfs by their $J-H$ and $H-K$ colors \citep{bessell,carpenter}.
M giants also dominate M dwarfs in catalogs of late-type stars to $K\sim$14.
These facts were used by \citet{maj03} to select M giants from 2MASS and map the streams of M giants from the Sagittarius (Sgr) dwarf spheroidal (dSph) galaxy and by \citet{sharma10} and \citet{rocha03,rocha04,rocha06} to map and track the Triangulum-Andromeda star cloud, the Pisces overdensity, and the Monoceros/GASS/Argo feature.
Our survey sample spans $(J-K_{S})_{0}$ colors from $0.75<(J-K_{S})_{0}<1.24$; this is similar to the range studied by \citet{girard06} in their study of the thick disk using red giants, although most (96\%) of our red giants have $(J-K_{S})_{0}>0.85$ (as in Majewski et al. 2003) to ensure a clean sample of M giants. 
The magnitude range of our entire sample is $4.3<(K_{S})_{0}<12.0$, with a median magnitude of 7.4 (the majority, 1625 of 1799, have $5.0<(K_{S})_{0}<9.0$).
The magnitudes were dereddened using the maps from \citet{schlegel98}.
    
\begin{figure}
\plotone{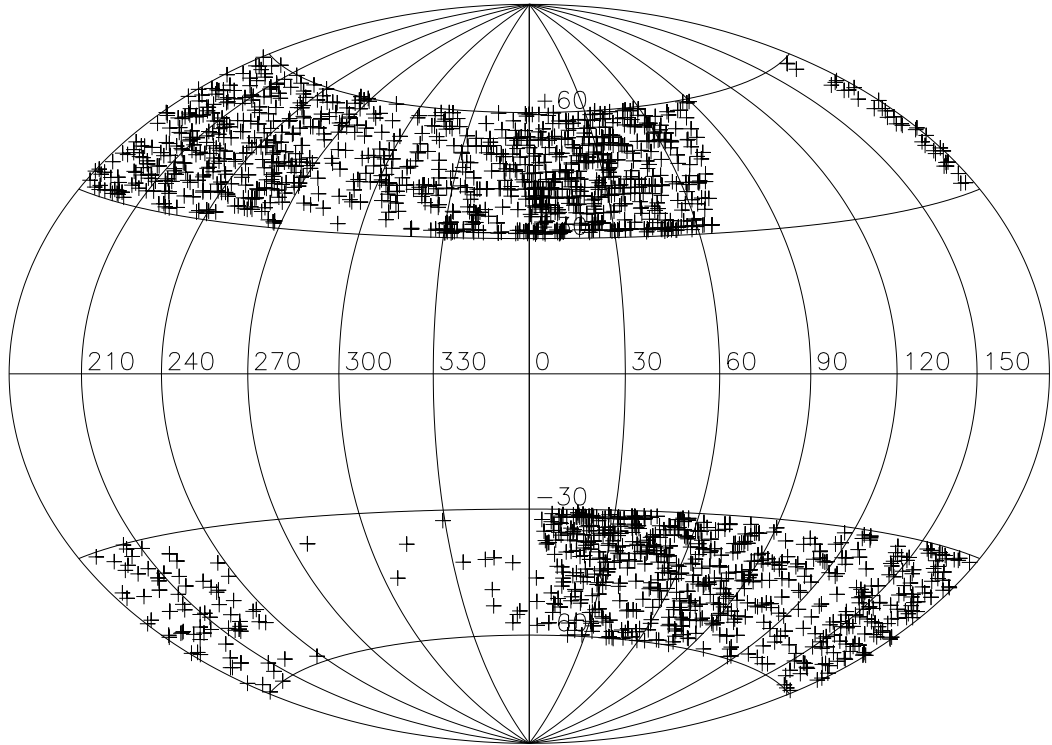}
\caption{\label{lb}Aitoff projection of the Galactic coordinates of the 1799 M giants observed at FMO and CTIO. The program began in February, 2005 and is ongoing.  Stars were selected to primarily sample a wide sweep of the Milky Way's thick disk and nearby halo.}
\end{figure}

A constraint in Galactic latitude of $30^{\circ}<|b|<60^{\circ}$ was applied to our M giant catalog, with the lower limit in $b$ applied to avoid excess contamination from the thin disk. 
Our nominal photometric sample contained approximately 12,000 stars, and in this paper we present spectroscopic observations for 1799, or roughly 15\% of these.
The stars observed were selected randomly from the nominal sample and cover nearly all Galactic longitudes. 
The bulk of the observing was done at the Fan Mountain Observatory (FMO), located in Virginia, so there are gaps in coverage corresponding to the Southern Hemisphere.  
Of the 1799 stars, 149 were observed at Cerro-Tololo Inter-American Observatory (CTIO) as part of a related program. 
Figure \ref{lb} shows in an Aitoff projection the spatial distribution of the 1799 program stars in Galactic coordinates.  
The nominal M giant catalog was matched to the UCAC2 catalog \citep{zacharias04} and the gap in the northern Galactic hemisphere is due to the upper limit of $\delta=+52^\circ$ in the UCAC2 catalog at the time of the survey inception.  
However, the very small amplitudes of the UCAC2 proper motions for the majority of the program stars typically result in unreasonably large relative errors (often larger than the derived space motions) in any derived kinematical parameters using them, so the proper motions are not actually utilized in the present work.  
Limitations from use of the UCAC2 proper motions on 2MASS M giants are further explored in \citet{mlpp}. 

\subsection{\label{fmo}FMO and CTIO Spectroscopic Observations and Reductions}
Spectra were collected at the University of Virginia's FMO using the Fan Observatory Bench Optical Spectrograph \citep[FOBOS; see][]{crane05} on the 1-m astrometric reflector.  FOBOS is a fiber-fed optical spectrograph that was designed for moderate resolution ($R\sim $ 1500-3000) spectroscopy \citep{crane05}.  
FOBOS uses a grating with 1200 grooves mm$^{-1}$.
The estimated resolution of the spectra is $\Delta\lambda\sim4$ \AA.  Our observing program began on UT 2005 February 25 and data presented here were taken in the years 2005 - 2008.  The spectrograph is optimized for use over the region 4000-6700 \AA\ but is limited by the camera, which has a SITe 2048$\times$2048 CCD detector that at a linear dispersion of 1.0 \AA\ pixel$^{-1}$ samples only $\sim$2000 \AA\ of that range (selected by us to be 4000-6000 \AA).  
On all but one night (when we were testing the efficiency of the set-up) the detector setting used had a read noise level of 4.5 e$^{-}$ and a gain of 6.1 e$^{-}$ per ADU. 
For most stars, three spectra are taken and summed in 2-D after the CCD frame preprocessing is completed.  This combination of three images facilitates the elimination of cosmic rays.  
The minimum S/N to achieve the best possible RV precision (a few km s$^{-1}$) was found to be $\sim$20; this typically translated to total exposure times of 450-900 seconds for the M giants observed with FOBOS, which have magnitudes in the range $4.3<(K_{S})_{0}<9.7$.
Several radial velocity standard stars from the Astronomical Almanac were also observed each night; these are used for cross-correlation templates when determining the radial velocities of the target stars.  
Standard stars were chosen to be of a similar spectral type as the program stars to minimize systematic offsets in the derived RVs.  Several sets of bias frames were taken throughout each night.  To remove pixel-to-pixel variations in the frames, a ``milky flat'' is created by illuminating an opal diffusing glass with a quartz-tungsten-halogen lamp.  The object and comparison frames are flat-fielded (using the milky flat), trimmed, and bias subtracted using the task $ccdproc$ in IRAF\footnote{IRAF (Image Reduction and Analysis Facility) is distributed by NOAO, which is operated by the Association of Universities for Research in Astronomy, Inc., under contract with the NSF.}.  Comparison spectra were taken using neon, argon and xenon lamps for calibrating wavelengths against the laboratory values for lines from these elements.  The spectra are extracted from the 2-D images and converted to 1-D spectra and wavelength calibrated using the IRAF tasks $apall$ and $identify$.  

To obtain radial velocities, the extracted, flat-fielded, wavelength-calibrated spectra are cross-correlated with the standard star spectra using code developed by W. Kunkel \citep[described in detail in][]{maj04b}.  The cross-correlation code is run for each night of data and each program star is cross-correlated to all of the standard stars (typically 4-6) observed on that night.  The radial velocity reported for a program star is the average of the radial velocities from cross-correlation with multiple standards taken that night (with standard star spectra that produce poor cross-correlations against the others removed from the average, iteratively). 
  
\begin{figure}
\plotone{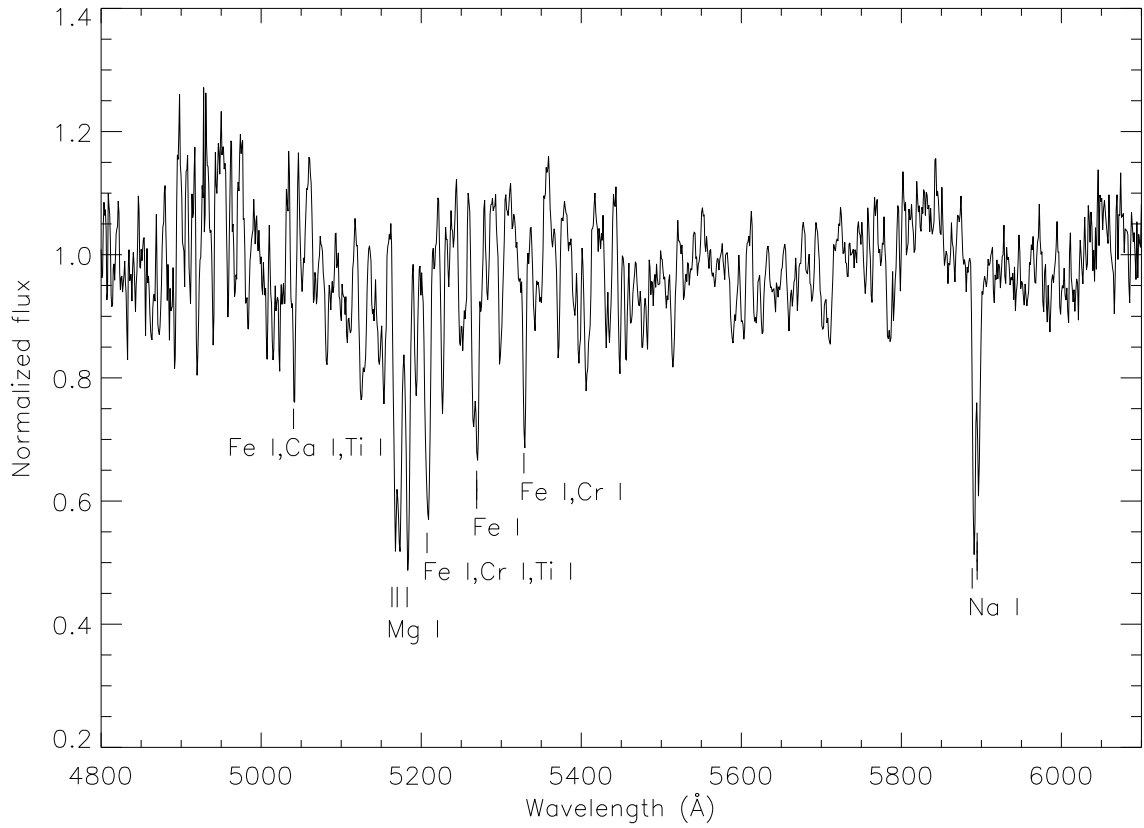}
\caption{\label{speclines}A typical medium-resolution spectrum for the program stars observed at FMO with FOBOS; prominent features --- the Mg b triplet, the Na D doublet, and several strong Fe, Cr, and Ti lines/blends --- are labeled.}
\end{figure}

\begin{figure}
\plotone{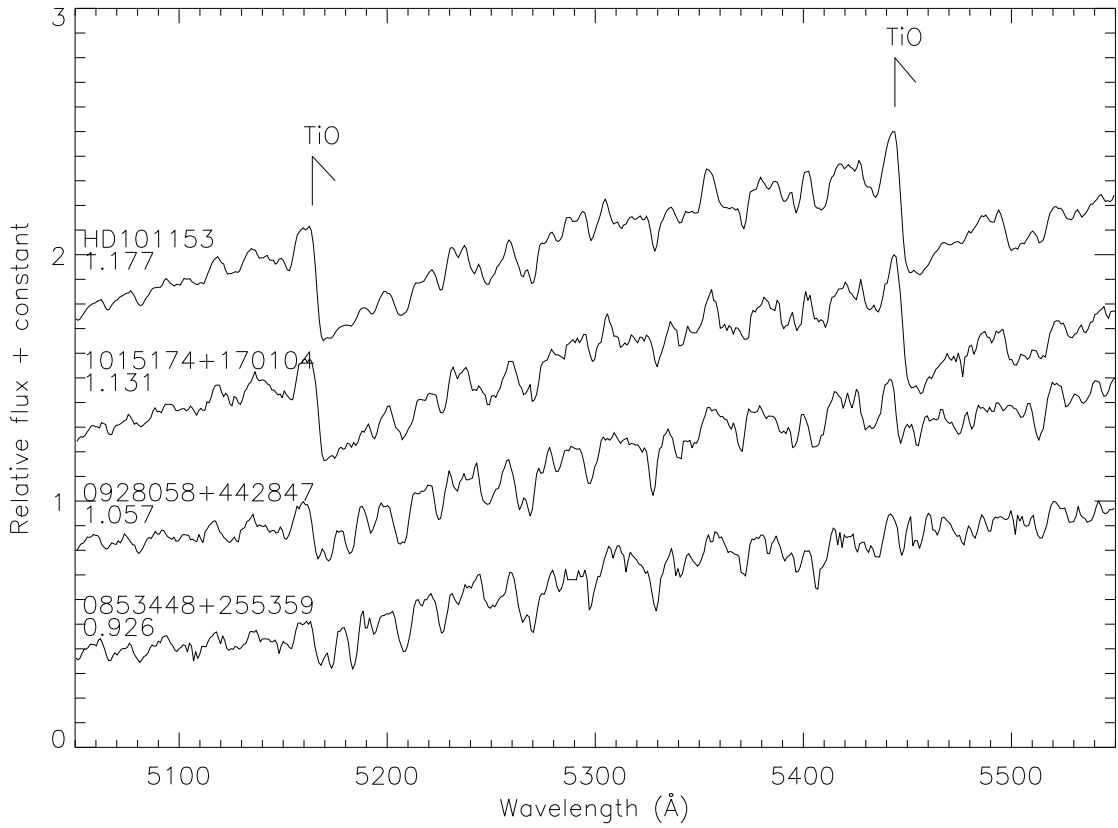}
\caption{\label{speccomp}Medium-resolution spectra of four stars observed at FMO with FOBOS, spanning a range of $(J-K_{S})_{0}$ colors for typical program stars.  The star IDs and colors are printed to the top-left of each spectrum.  The topmost spectrum is the type M4 giant HD101153.  The increasing prominence of molecular bands is apparent for the redder (cooler) stars.}  
\end{figure}

The spectrum for a typical program star observed with FOBOS is shown in Figure \ref{speclines}. 
The dominant features in M giants in the spectral band we are studying are the Mg b triplet, at 5167, 5173, and 5184 \AA, and the Na D doublet, at 5889 and 5896 \AA.  A number of strong Fe, Cr, and Ti lines/blends are also present.  The redder stars in our sample ($J-K_{S}>1.1$) show very strong titanium oxide (TiO) bands in their spectra; these M giants are cool enough ($T_{\rm eff}<3560$ K) that the TiO bonds in the star are not dissociated.  Figure \ref{speccomp} shows the spectra for several stars observed with FOBOS covering a range of ($J-K_{S})$ colors; the strength of the TiO bands increases as the giants become redder and cooler.     

An additional 149 M giants that fit our survey criteria were observed over UT 2004 October 8 - 11 at CTIO using the Cassegrain spectrograph on the 1.5-m telescope.  The detector was a Loral 1K (1200$\times$800 pixels) CCD with a read noise of 6.5 e$^{-}$; the gain was set to 1.42 e$^{-}$ per ADU.  
A grating with 831 grooves mm$^{-1}$ was used, with a resolution of $\Delta\lambda\sim3.1$ \AA.
Helium and argon lamps were used to take comparison frames for each target at the same telescope position to account for flexure variations.  Ten quartz frames were taken each night and combined and normalized.  The spectral range is 7650-8900 \AA.  
The dominant feature in this region is the Ca II triplet at 8498, 8542, and 8662 \AA.
The CTIO M giants have magnitudes in the range $6.3<(K_{S})_{0}<12.0$.
The reduction procedures for the CTIO program giants are similar to those used for the FMO reductions (the same cross-correlation code was used to determine the radial velocities). 

For the FMO sample, the difference in the derived (from cross-correlation against each other) and published RVs for the standard stars is typically $\pm$1-5 km s$^{-1}$, with no systematic offset in either direction.  
A total of 102 program stars were observed multiple times at FMO to test the stability of FOBOS and to gauge the S/N threshold for obtaining reliable RVs. 
The mean of the absolute value of the differences in the RVs for stars with multiple observations at FMO is 8.5 km s$^{-1}$.
The RVs are fairly stable even at low S/N:
stars with S/N below 20 have a mean absolute value in the difference of their RVs of 10.5 km s$^{-1}$. 
For the CTIO sample, the difference in the derived and published RVs for the standard stars is slightly higher, with variations ranging from $\pm$1-10 km s$^{-1}$; as with the FMO standards, no systematic offset in either direction is seen.  
Repeat observations were also taken for 16 stars at CTIO.
The mean of the absolute value of the differences in RVs for the CTIO repeat observations is 13.9 km s$^{-1}$. 
For a subsample of 34 stars, radial velocities were also determined from high-resolution echelle spectra (see \S\ref{hirvs}).
In Table \ref{tab2}, the radial velocities found from the high- and medium-resolution spectra for these 34 stars are reported (high-resolution/medium-resolution, denoted as $v_{\rm hel,h}/v_{\rm hel,m}$, respectively). 
The mean of the absolute value in the differences between the medium-resolution RVs with the high-resolution RVs is 6.2 km s$^{-1}$.
Overall, considering the random errors in the medium-resolution RVs, the comparison of the medium/high resolution values, and that the data set is dominated by FOBOS observations, we place the typical uncertainty level for the medium-resolution RVs at 5-10 km s$^{-1}$.

\section{Radial Velocity Distribution\label{rvs}}
\begin{figure}
\begin{center}
\epsscale{1.0}
\plotone{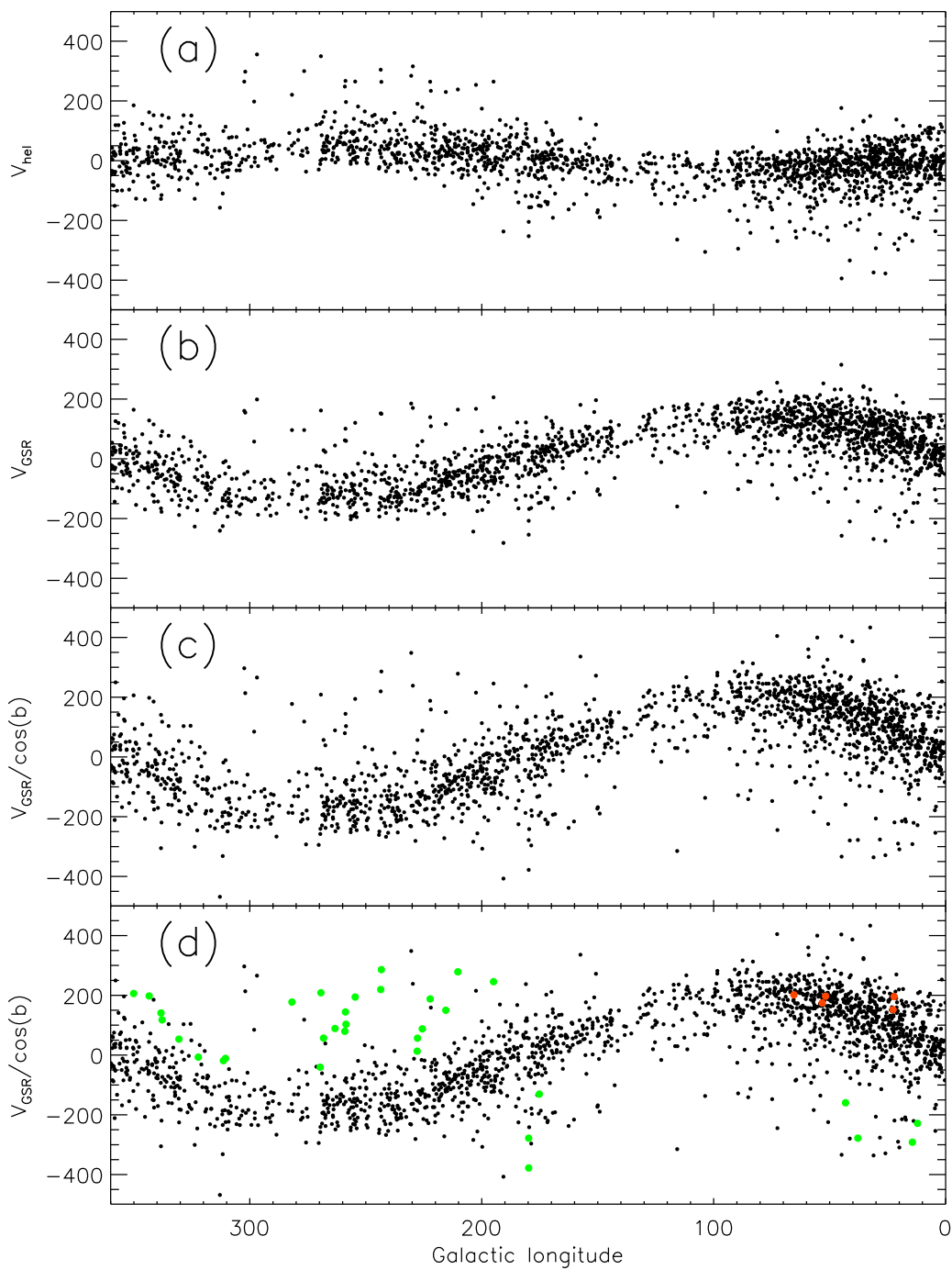}
\caption{\label{rvcosb_3pan}Radial velocities as a function of Galactic longitude for all 1799 program M giants. Panel (a) shows $v_{\rm hel}$ vs. $l$ and panel (b) shows $v_{\rm GSR}$ vs. $l$.  In panel (c), $v_{\rm GSR}$ has been divided by $\cos(b)$.  Panel (d) repeats panel (c), where stars with high-resolution spectra obtained are overplotted in color -- green for ``RV outliers'' and orange for five stars with RVs that match those of the bulk thick disk trend.}
\end{center}
\end{figure}

Panel (a) of Figure \ref{rvcosb_3pan} shows the heliocentric radial velocity, $v_{\rm hel}$, as a function of Galactic longitude, $l$.\footnote{The full medium-resolution radial velocity catalog is available upon request.} 
In panel (b), these velocities have been translated to the GSR (Galactic standard of rest frame --- i.e., centered on the Sun but at rest with respect to the Galactic Center), where we adopt the values $\Theta_{0}$=236 km s$^{-1}$ for the speed of a closed orbit at the position of the Sun relative to the Galactic center \citep{bovy09} and ($U_\odot$,$V_\odot$,$W_\odot$)=(11.10,12.24,7.25) km s$^{-1}$ \citep{schonrich10} for the motion of the Sun with respect to this orbit.  
General trends in these panels can be understood by assuming that stars in the Galactic disk move on nearly circular orbits around the Galactic center. 
For a flat Milky Way rotation curve near the Sun with circular speed $\Theta_{0}$, the predicted line-of-sight velocity with respect to the GSR at the Sun's position for a star on a circular orbit is given by:
\begin{equation}v_{\rm GSR,circ}= \Theta_{0} \left(\frac{R_0}{R}\right) \sin   l \cos b \label{vpred} \end{equation}
Here, $R$ is the Galactocentric radius of the star and $R_0$ is the solar Galactocentric radius.
The expected sinusoidal trend with $l$ for stars moving on disk-like orbits around the Galactic center is seen for the bulk of the stars.
Note that the heliocentric velocities in panel (a) still show some sinusoidal trend --- a reflection of the asymmetric drift of the M giant population (i.e., the tendency of older stars to have circular velocities that lag the Local Standard of Rest); this is discussed further in \S\ref{interp}.

At fixed $l$, equation (\ref{vpred}) shows that $v_{\rm GSR,circ}$ is lower for stars on circular orbits observed at high $b$ than for stars observed at low $b$ due to the $\cos b$ projection of their motions.  
In principle, therefore, one may have a good approximation to the rotational component of a disk star's velocity at a latitude $b$ by deprojecting the observed velocity with a division by $\cos b$.
This scaling tends to accentuate differences between stars having ``disk-like'' (i.e., circular) motion from stars having non-disk-like motions
because the deprojection will tend to tighten the coherence of disk stars but separate outliers more --- as shown in panel (c) of Figure \ref{rvcosb_3pan} and in Figure 1 of \citet{maj12}.

While the majority of the stars in Figure \ref{rvcosb_3pan} appear to be members of the thick disk based on the amplitude of the sinusoidal trend in panel (a), there are a significant number of stars with high velocities relative to the main trend (we refer to these stars as ``RV outliers'').
The origin of these RV outliers --- {\it in-situ-halo, kicked-out}, or {\it accreted} --- is unclear from this observational plane alone. 
If the RV outliers are \emph{in-situ-halo} or \emph{kicked-out} stars, we would expect to see random RVs as a function of $l$. 
However, stars in several longitude ranges in our sample show suggestive coherence in their RVs.
Such coherence is expected for a stream of stars passing through the Solar neighborhood \citep[see][for further details]{maj12,johnston12}, but  
whether these structures are real is hard to assess given the small number of stars.

In the next section we present high-resolution follow-up spectroscopy of a sample of 34 RV outliers (highlighted by green symbols in the lower panel of Figure \ref{rvcosb_3pan}) to examine what fraction of outliers can be attributed to each of the halo formation mechanisms.

\section{Chemical Abundances\label{hires}}

\subsection{Data Collection and Reductions\label{hirvs}}
To test the chemical properties of the 34 selected stars,
high-resolution echelle spectra were collected using the Astrophysical Research Consortium Echelle Spectrograph (ARCES) on the 3.5-m telescope at Apache Point Observatory on UT 2009 March 30 and UT 2009 April 2 and the CCD echelle spectrograph (ECHLR) on the 4-m Mayall telescope at the Kitt Peak National Observatory over UT 2010 February 26 - March 2.  

The ARCES uses a 2048$\times$2048 pixel SITe CCD and has a resolution of $R$=31,500; the CCD has a gain of 3.8 e$^{-}$ per ADU and a readout noise level of about 7 e$^{-}$.  
Two sets of flat fields were taken to account for the strong gradient in response across the ARCES orders: one long set with a blue filter inserted and another short set with no filter (these are the red calibration frames).  The blue and red frames were combined to create a master quartz flat.  For wavelength calibration, thorium-argon (ThAr) lamp frames were taken.

Reduction of the ARCES data was carried out using various IRAF tasks in the $echelle$ package.  All images were overscan-corrected and trimmed using $ccdproc$.  The echelle orders were located and the trace defined for the spectra with $apall$.  The task $ecidentify$ was used to identify lines in a ThAr lamp spectrum. To minimize the effects of aliasing, the spectra were resampled along the dispersion axis using the $magnify$ task.  Scattered light was removed from the program star frames using the $apscatter$ task, and the relevant echelle orders were then extracted.  The program spectra were divided by the extracted master quartz flat, and the dispersion correction defined using the ThAr lamps was then applied to convert the pixel scale to a wavelength scale.  As a final step in the reduction process, the spectra were continuum normalized. 

ECHLR spectra were collected at KPNO using the 2048$\times$2048 pixel T2KB CCD on the 4-m Mayall.  The gain setting for T2KB was 1.9 e$^{-}$ per ADU with a read noise of approximately 4 e$^{-}$. The ECHLR has a resolution of $R$=35,000.  
ThAr lamps were used to collect wavelength calibration frames, and the short-exposure quartz lamp was used for obtaining the flat-field images.  The spectral reduction procedures for the ECHLR data are similar to those carried out for the ARCES data. 

The photometric properties and the observational details of the stars observed at APO and KPNO are listed in Table \ref{tab2}.  
The IDs and photometry of the stars come from the 2MASS point source catalog, where the IDs are the 2MASS RA/Dec (J2000.0) coordinates.
Radial velocity standards, taken from the Astronomical Almanac, were observed all nights at APO and KPNO.  
Radial velocities from the echelle data were found using the IRAF $fxcor$ task and are reported in Table \ref{tab2} and are compared with the medium-resolution values. 
Cross-correlation between RV standards gives errors on the order of 0.5 km s$^{-1}$ for the high-resolution RVs.
Based on the RV data in Table \ref{tab2}, there is apparently a systematic bias towards higher measured RVs for the medium-resolution spectra, such that $v_{\rm hel,h}-v_{\rm hel,m}$=-5.0 km s$^{-1}$.   
This offset may be due to variations in the centering of the star in the slit for the echelle data (FOBOS data are taken with fiber optics, which, due to radial and azimuthal scrambling, provide more uniform ``slit functions'' in the spectrograph).

\subsection{\label{abund}Chemical Abundances}
\subsubsection{Derivations}
The APO and KPNO instrument set-ups give the best S/N per pixel for the region around 7400 \AA.  This particular region of the spectrum was selected due to its relative absence of molecular bands, which offers a good window for spectral analysis \citep{sl90}.  
The procedures used to derive the metallicities are similar to those used by \citet{chou07} in their study of M giants in the Sgr tidal tails.  
Equivalent widths (EWs) for 11 Fe I lines in the range 7443 \AA\ to 7583 \AA\ were used to determine the atmospheric parameters $T_{\rm eff}$, log $g$, and [Fe/H] and the microturbulent velocity ($\xi$).  
The EWs were measured manually using the IRAF $splot$ task.

\begin{figure}
\epsscale{1.0}
\plotone{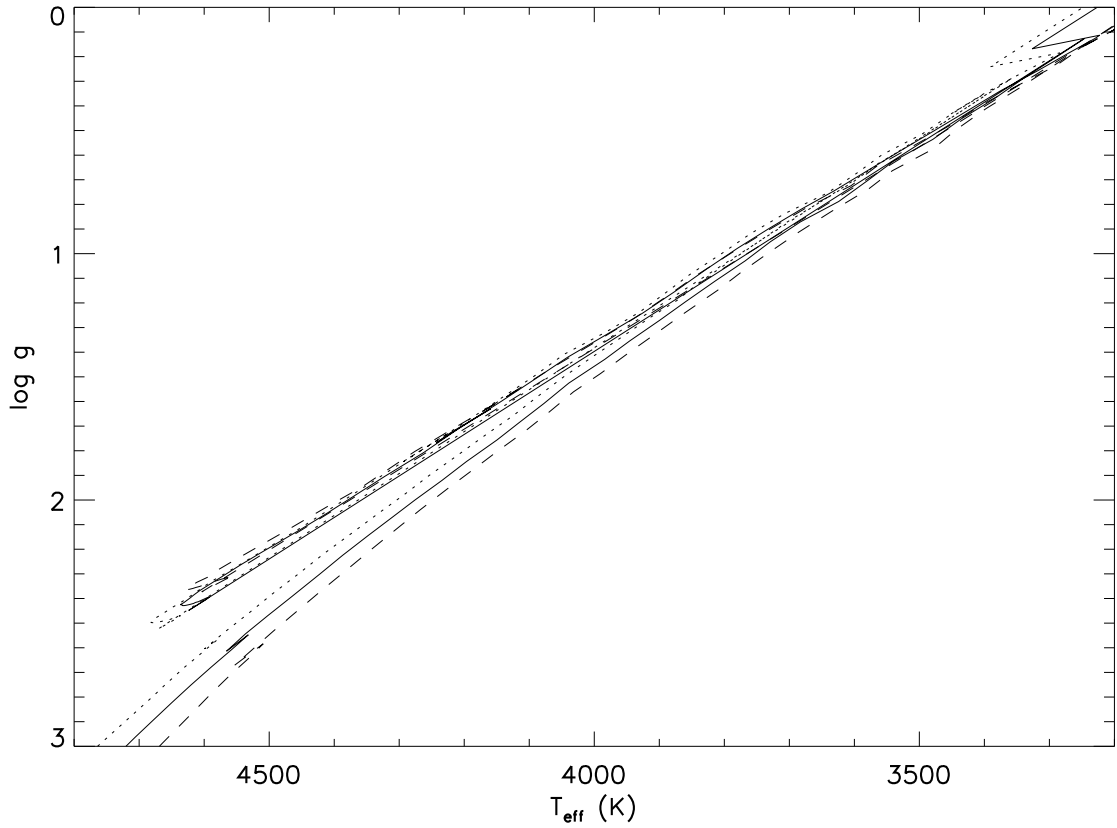}
\caption{\label{teff}$T_{\rm eff}$-log $g$ isochrones \citep{marigo08,girardi10} for populations of age 3 Gyr (dotted line), 5 Gyr (solid line), and 10 Gyr (dashed) with $Z=Z_{\sun}=0.019$.  The range of $T_{\rm eff}$ derived for the APO/KPNO program stars is 3600-4000 K. At higher $T_{\rm eff}$, the log $g$ values begin to diverge significantly.}  
\end{figure}

\begin{figure}
\plotone{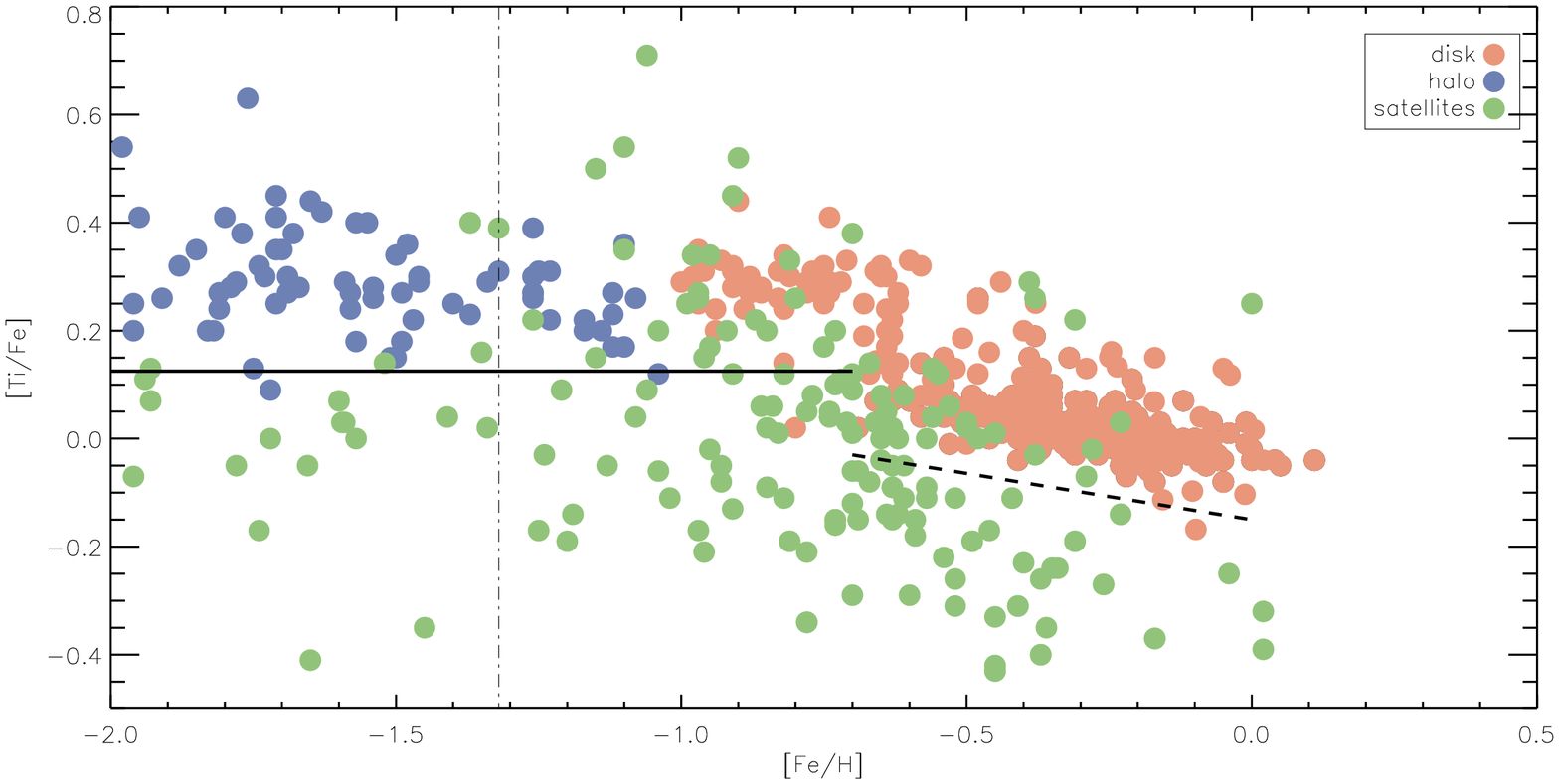}
\caption{\label{tag}[Ti/Fe] vs. [Fe/H] for stars from the literature color-coded as: orange (disk), blue (halo) and green (satellite).  Disk stars are from \citet{fulbright00}, \citet{reddy03}, and \citet{brewer06}, where stars from the \citet{fulbright00} sample with [Fe/H]$>$-1 are categorized as disk stars and those with [Fe/H]$<$-1 are categorized as halo stars. Satellite stars are from Milky Way dSphs (Carina, Fornax, Sagittarius, Sculptor, and Sextans) and red giants in the LMC.  The dSph data are from \citet{shetrone01,shetrone03}, \citet{monaco05}, and \citet{chou07} and the LMC data come from \citet{smith02} and \citet{pompeia08}.  Stars to the left of the dashed diagonal line and below the solid horizontal line -- determined by eye -- separate a nearly pure population of stars from Milky Way satellites. 
The dot-dashed line at [Fe/H]=-1.3 marks the minimum metallicity for our high-resolution red giant sample.}  
\end{figure}

The wavelengths of the 11 Fe I lines and their corresponding EW measured for each star are listed in Table \ref{tab3}.  In some cases, an EW could not be measured due to a cosmic ray falling on the same pixel as an Fe line; these lines were removed from the list for that spectrum and are reported as ``...'' in Table \ref{tab3}.  We adopt the same values for the excitation potentials ($\chi$) and oscillator strengths ($gf$) for these lines as those used by \citet{chou07}.  Along with the EW measurements, stellar atmosphere models from the Kurucz ATLAS9 grids (1994) are used.  
Each stellar atmosphere model corresponds to a particular ($T_{\rm eff}$, log $g$, [Fe/H]).  
We start with an initial guess for ($T_{\rm eff}$, log $g$, [Fe/H], $\xi$)$_{0}$.  
The EWs for the sample lines and the model atmosphere are used as input to the MOOG local thermodynamic equilibrium (LTE) code \citep{sneden73}.
The starting value for $T_{\rm eff}$ is calculated from the \citet{houdashelt00} color-temperature relations for M giants, using the dereddened 2MASS $J-K_{S}$ colors converted to the CTIO system and the relations of \citet{carpenter}.  An initial guess for log $g$ is taken from the 10 Gyr $Z_{\sun}$ isochrone from the Padova evolutionary track database (Marigo et al. 2008\footnote{see http://pleiadi.pd.astro.it/}).  The surface gravity $g$ of a star is related to its mass and size (luminosity).  Once a star exhausts the hydrogen supply in its core, it will first brighten and move up the red giant branch (RGB); later, after core He exhaustion, the star eventually ascends the asymptotic giant branch (AGB).  The difference in log $g$ between a star on the RGB vs. the AGB for cool giants, however, is quite small over a wide range of ages.  
Figure \ref{teff}, which shows the $T_{\rm eff}$-log $g$ plane for 3 Gyr, 5 Gyr, and 10 Gyr solar metallicity ($Z_{\sun}$=0.019) isochrones, demonstrates that the variation in log $g$ between the RGB and AGB is on the order of 0.15 dex for stars in the range of effective temperatures for our sample, which is 3600-4000 K.  
The iterative scheme for determining abundances involves using the initial values of $T_{\rm eff}$, log $g$, and [Fe/H] and iterating these values until the derived and model values of [Fe/H] converge.
After each iteration, the correlation coefficient for $A$(Fe) as a function of the reduced EW (i.e., RW -- the measured EW divided by the $\lambda$ required for that atomic transition) is checked and $\xi$ is adjusted to minimize the correlation coefficient before proceeding to the next iteration.  
An incorrect value of $\xi$ leads to a physically unrealistic dependence of the elemental abundance on the EWs -- as seen by a high correlation coefficient for the RW.  
The derived parameters for all program stars are listed in Table \ref{tab4}.

In addition to [Fe/H], the ratio [Ti/Fe] was also derived. Once the atmospheric parameters for a star were determined using MOOG, these were fixed and used to measure $A$(Ti). Three Ti I lines were used to derive $A$(Ti): 7474.940, 7489.572, and 7496.120 \AA.
A solar $gf$ value was derived for the Ti I line at 7474.94 \AA\ and adopting $A$(Ti)$_{\sun}$ = 4.90 from \citet{asplund05}. The $gf$ values for the other two Ti I lines (7489.57 and 7496.12 \AA) were taken from \citet{chou07} and these are also solar $gf$ values.

As a check of our methodologies, four calibration stars from the red giant spectral studies of \citet{sl85} and \citet{smith00} (collectively referred to as S\&L) were observed with both the ARCES and ECHLR; a comparison between our [Fe/H] and [Ti/Fe] values for the calibration stars is given in Table \ref{SLcomp}.  
Typical uncertainties in the values derived for [Fe/H] in both studies are $\sim \pm$0.15 dex using this sample of Fe I lines.  The mean difference in [Fe/H], and its associated standard deviation, between these two studies, in the sense of (this study - S\&L) = -0.04 $\pm$ 0.16.
This comparison indicates that the two studies are on the same abundance scale, with only a small mean offset and a scatter typical of the derived uncertainties in [Fe/H]. 
The values are in good agreement and the discrepancies are due to differences in the stellar atmosphere models used. 

Although detailed non-LTE (NLTE) calculations have not been carried
out here, limits to the simplifying assumption of LTE can be
investigated using recent studies of NLTE line formation for iron
by \citet{bergemann11b} and for titanium by \citet{bergemann11a}.
As is the case for many NLTE calculations involving cool stars, the
theoretical results depend on the choice of the value for the
parameterized efficiency of neutral hydrogen collisions, S$_{\rm H}$.
In the case of Fe I/Fe II, \citet{bergemann11b}
computed NLTE/LTE results using a small grid of model atmospheres and
find that for Fe I in general, corrections to LTE-based abundances
will become larger for increasing $T_{\rm eff}$, decreasing surface
gravity, or decreasing metallicity; the Fe I corrections increase most
dramatically for decreasing [Fe/H] (see their Figure 3), especially for
[Fe/H]$\le$ -2.0, and for warmer temperatures, $T_{\rm eff}$$\ge$5000 K
(where $\Delta$(NLTE -- LTE)$\ge$+0.2 dex).  The sample of M-giants
analyzed here is only moderately metal-poor ([Fe/H]$\ge$-1.3) and, when
coupled to cooler effective temperatures, would be expected, based on
the \citet{bergemann11b} results, to require corrections to LTE Fe I
abundances that would be $\le$+0.2 dex.  Although NLTE corrections to
LTE abundances in cool giants are not expected to be large, NLTE
theoretical calculation of corrections that span the stellar parameter space analyzed here
would be welcome.  

The Ti I corrections to LTE from \citet{bergemann11a} are in the same sense
and qualitatively similar to those for Fe I discussed above.  Given the
stellar parameters and metallicities covered by the sample here, any
corrections to [Ti/H] are not significant and, in particular, the
critical values of [Ti/Fe] will have even smaller corrections from
assuming LTE, with values $\le$0.10 dex; such uncertainties have no
significant effect on conclusions drawn from the Fe and Ti abundances calculated here.
As noted in \citet{bergemann11b}, their goal is to establish
interactive routines to allow for estimating corrections to LTE-based
abundances, so in the near future, it may be possible to provide more
accurate corrections to LTE given particular stellar parameters and
metallicities.

Observations of stars in clusters also support the theoretical
calculations that suggest rather small departures from LTE for the
Fe abundances in giants.  \citet{ramirez01} analyzed stars in
the mildly metal-poor globular cluster M71 ([Fe/H]$\sim$-0.5) and
found the same values of [Fe/H] (within $\sim$0.05 dex) for turn-off
stars, subgiants, and giants (which span a range in $T_{\rm eff}$ from
6000 K to 4500 K and log $g$ from 4.1 to 1.5).  These results support
the small-ish corrections to LTE Fe abundances suggested by the
\citet{bergemann11b} calculations.

\subsubsection{Distances}
High-resolution spectra allow us to compute distances.
Although we don't use them explicitly in most of our analysis, these distances, reported in the last column of Table \ref{tab4}, are helpful to obtain some understanding of the size of the volume our M giant sample probes around the Sun. 
The distances to the high-resolution program stars are computed using isochones from the Padova database \citep{marigo08,girardi10}.  
Approximate distances are determined using the derived [Fe/H] and assuming stars of age 10 Gyr; the appropriate $M_{K}$ is then selected based on the derived $T_{\rm eff}$.
Using an age of 10 Gyr means that a star evolving along the RGB or AGB would have had an initial mass of $\sim$ 1.0 $M_{\sun}$; such a typical mass would not be largely different from a much younger population, such as 2 Gyr where $M_{RGB/AGB}\sim$ 1.7$M_{\sun}$, or a population with an age of 5 Gyr and $M_{RGB/AGB}\sim$ 1.1-1.3$M_{\sun}$. 
As discussed above (see Figure \ref{teff}), the stellar atmospheric parameters for a red giant on the RGB do not vary significantly from those for a red giant on the AGB. 

The derived distances of the standards agree well with their Hipparcos distances (10\% to 14\% accuracy) when using isochrones in the same way as used for the high-resolution program stars.
Possible sources of systematic error in the derived distances include: (1) an incorrect assumption of the star's age and (2) scatter in the isochrone for later evolutionary stages. 
As a check on the error in distance from the isochrone method, $M_{K}$ was computed for ages of 5, 8, and 10 Gyr for stars having different values of $T_{\rm eff}$ and [Fe/H].
Using estimated uncertainties in $T_{\rm eff}$ of $\sim$100 K and $\sim$0.10 dex in [Fe/H], a conservative distance uncertainty (from the variation in $M_{K}$ with age, $T_{\rm eff}$, and [Fe/H]) is $\sim$25\%, with the scatter in the distribution of distances expected to be somewhat smaller than this.

We note that a handful of the more metal-poor stars cannot be fit to a 10 Gyr isochrone because their values of $T_{\rm eff}$ are cooler than the RGB tip.
It is possible that this subset of stars consists of cool AGB stars, as the low-metallicity, low-mass stellar models do not model the end of the AGB well.

Based on the errors in the derived distances for the standards and the sources mentioned above, we assign an approximate accuracy of 20\% to the distances for the program M giants. 
The distances to all high-resolution program stars range from 0.9 to 8.9 kpc with a mean distance of 4.3 kpc; these distances confirm that the survey sample probes out to as far as roughly 9 kpc from the Sun, and, in the mean, sample a good volume of the near side of the Galaxy thick disk and halo components around the Sun. 

\section{\label{interp} Results \& Interpretation} 

Figure \ref{tag} empirically justifies the locations for the blue, green and orange regions that were sketched in Figure \ref{cartoon} by plotting observed abundances for samples of halo (blue symbols), Milky Way satellite (green symbols), and disk (orange symbols) stars.
As expected there is significant overlap between these populations.
Nevertheless, the dark horizontal and dashed diagonal lines in this plot quite clearly separate an almost pure sample of satellite stars. 
Hence, we can assign possible origins of stars in our sample based on their location in the [$\alpha$/Fe]-[Fe/H] plane as follows: (i) stars below the horizontal line and to the left of the diagonal line are very likely to have been {\it accreted};
(ii) stars to the right of the diagonal line along the main disk trend 
may have formed deep within the dark matter halo of the Galactic progenitor, but could also have been accreted from a high mass infalling object (like the LMC or Sgr, which have stars that fall in this region);
(iii) stars above the horizontal line along the main halo trend could have been accreted or could have formed early in the life of the Galaxy in either the disk or the halo.

\begin{figure}
\plotone{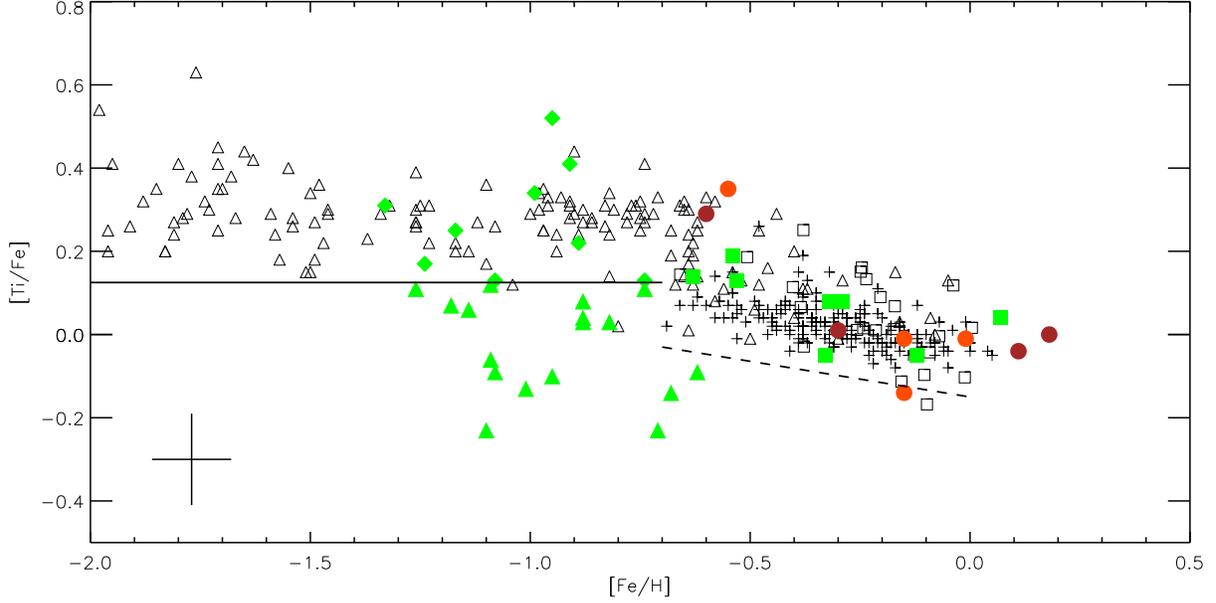}
\caption{\label{tife}[Ti/Fe] vs. [Fe/H] for program stars with high-resolution spectra.
The high-resolution M giants are plotted in green and have been coded as follows: filled triangles correspond to {\it accreted} stars; filled squares correspond to {\it kicked-out} stars; filled diamonds could be {\it accreted}, {\it kicked-out}, or {\it in-situ-halo} stars.  
Milky Way disk and field stars taken from the literature are shown in black and are coded as follows: open triangles - \citet{fulbright00}; plus signs - \citet{reddy03}; open squares - \citet{brewer06}.
As a control for the red giant standard stars, four thick disk stars (those with origin labeled as `d' in Table \ref{tab2}) -- based on RVs from Fig. \ref{rvcosb_3pan} -- are shown as orange circles (the star 1535178+135331 is not shown due to a large error in its derived [Ti/Fe]). The four red giant standards Arcturus, $\alpha$ Tau, $\nu$ Vir, and $\delta$ Vir shown as brown circles.  Typical error bars are shown in the lower left in black.}  
\end{figure}

Figure \ref{tife}, with the horizontal and diagonal lines repeated from Figure \ref{tag}, shows [Fe/H] vs. [Ti/Fe] for the 34 program M giants and 8 of the 9 calibration stars having high-resolution spectra (one calibration star, 1535178+135331, is not included due to a high uncertainty in its [Ti/Fe]), along with the literature data for Milky Way field stars. 
  
The locations of our RV outliers in Figure \ref{tife} can be interpreted in terms of stellar halo formation scenarios:
\begin{itemize}
\item
Seventeen RV outliers (shown as filled green triangles) fall outside both the disk or halo chemical trends. 
This suggests a lower limit of 17 {\it accreted} stars in our sample. 
\item 
Nine of our RV outliers fall along the main Galactic halo trend (above the solid line, and shown as filled green diamonds).
Because of the expected chemical overlap in formation scenarios, the origin of stars in this region is ambiguous --- they could be \emph{in-situ-halo}, {\it kicked-out} or {\it accreted}.
However, the presence of 17 stars in the {\it accreted} region of the abundance plane, which is populated only during the later stages of chemical evolution in a system, suggests that there also must be some stars {\it accreted} from the same systems in the main Galactic halo region.
This implies that there are less than 9 {\it in-situ-halo} stars in our sample.
\item 
Eight RV outliers fall along the high-metallicity Galactic disk trend (to the right of the dotted line and shown as filled green squares). 
The combination of disk-like chemistry but extreme kinematics for these stars suggest membership of the {\it kicked-out} population. 
However, they could be a contaminating contribution from either: (i) a high mass accreted satellite (like Sgr or the LMC); or (ii) the true Galactic disk.  
\end{itemize}
These tentative population assignments for the program M giants are given in the last column of Table 1, where `is' refers to \emph{in-situ-halo}, `ko' refers to {\it kicked-out}, and `a' refers to the \emph{accreted} population.  The five stars with an origin of `d' are red giant thick disk calibration stars, selected based on the fact that their medium-resolution RVs fall along the main thick disk RV trend in Figure \ref{rvcosb_3pan}. 

\begin{figure}
\plotone{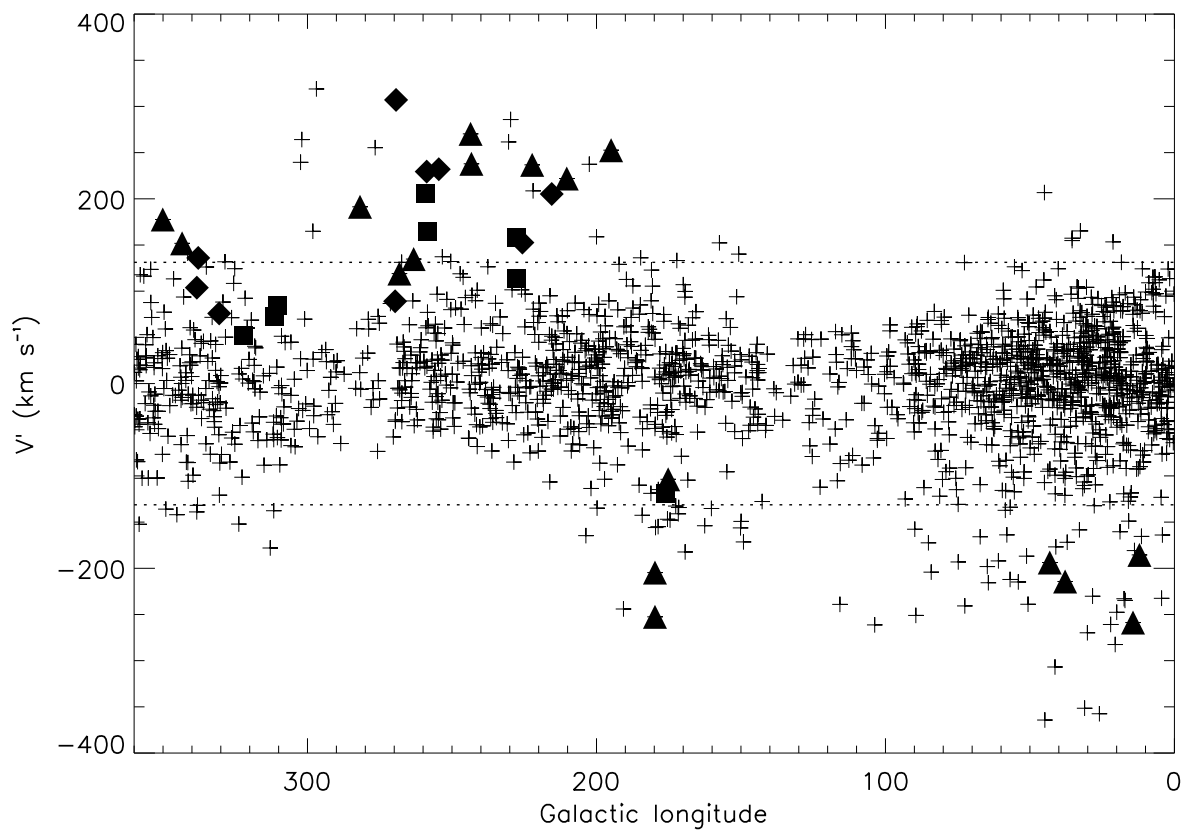}
\caption{\label{vasymm} $V^{\prime}$ as a function of Galactic latitude for the 1799 medium-resolution program stars (plus signs), with the 34 program stars with high-resolution spectra over-plotted using filled symbols (same population classifications as Figure \ref{tife}).  The dotted lines show 2.5$\sigma_{RV,disk}$.  The velocity $V^{\prime}=v_{\rm hel}+v_{\rm asymm}\sin(l)\cos(b)$ is discussed in \S\ref{interp}.}  
\end{figure}

We can derive a crude upper-limit to the contamination of our possible ``kicked-out'' stars due to true Galactic disk members by simply asking what fraction of thick disk stars could be moving at high enough speeds to be in our RV outlier sample.
\begin{enumerate}
\item
We characterize the motions of the disk population in our M giant sample 
by finding the value of the asymmetric drift $v_{\rm asymm}$ (apparent in the sinusoidal trend in the top panel of Figure \ref{rvcosb_3pan}) that minimizes the dispersion $\sigma^{\prime}$ of $V^{\prime}=v_{\rm hel}+v_{\rm asymm}\sin(l)\cos(b)$ 
calculated (iteratively, using a 3-$\sigma$ clipping method to remove outliers) for the medium-resolution heliocentric radial velocities.
Figure \ref{vasymm} shows $V^\prime$ for $v_{\rm asymm}=55$ km s$^{-1}$, which was found to give the minimum  $\sigma^{\prime}=52.5$ km s$^{-1}$.
\item
If {\it all} 1799 M giants were disk members
we would expect 1.2\%, or 22 stars, to have $|V^{\prime}| > 2.5\sigma^{\prime}$ (or 131.3 km s$^{-1}$).
In fact, there are 113 stars in our medium-resolution sample with such high $V^{\prime}$, which suggests that the majority of these are members of another population (i.e., the stellar halo). 
\item
This suggests that 22/113=19.5\% is a rough upper bound to the fraction of our RV outliers that are true Galactic disk members.
Since there are 24 stars in our high-resolution sample with $|V^{\prime}|$ $>$ 131.3 km s$^{-1}$, we estimate at most 5 of these could be true disk members.
\end{enumerate}
We actually find that 3 out of the 8 possible {\it kicked-out} stars have  $|V^{\prime}|$ $>$ 2.5 $\sigma^{\prime}$ (with 2 more falling on the border, as seen in Figure \ref{vasymm}); thus, we cannot claim any conclusive evidence of kicked-out stars in our high-resolution RV outlier sample.

Overall, we conclude that RV outliers in our M giant sample appear to be dominated by  {\it accreted} stars (more than 17), 
with a possible contribution from {\it in situ} stars.

\section{\label{conc}Summary, Discussion and Conclusion}

\subsection{Key Results}
In this study, we have analyzed the spectroscopic properties of a sample of M giants that are dominated by members of the Milky Way's thick disk and nearby halo components.
From the RV distribution, we identified stars with RVs that lie outside those expected for typical thick disk stars at the same locations.
These RV outliers are found to show some degree of spatial and kinematical coherence.   
We suppose that some of this coherence could be the signature of substructure (e.g., tidal tails) from accreted satellites.

To test our interpretation of the RV outliers, we looked at the chemical abundance patterns of 34 of these stars.
We used the locations of the stars in the [Ti/Fe] vs. [Fe/H] plane to attempt to assign them to one of three potential populations of halo stars -- those formed {\it in situ}, those that were {\it accreted}, and those that were {\it kicked out} of the disk.
This cannot be done unambiguously in all cases because of expected overlaps in the chemical signatures of these populations.
However, the [Ti/Fe] abundances for seventeen of the RV outliers are systematically below the main halo trend and
similar to the abundances seen in stars from Milky Way dwarf satellite galaxies.
They are consistent with stars from MW dwarf satellite galaxies that have been {\it accreted} by the Milky Way \citep{shetrone03,tolstoy04,monaco07} and inconsistent with expectations for abundance patterns from pure {\it in-situ-halo} or {\it kicked-out} growth. 
Another eight of the stars selected for high-resolution spectroscopic follow-up track the abundance trends set by the disk stars, even though they have halo-like kinematics.
These are indicative of a population formed in the disk or bulge and subsequently {\it kicked-out}, although: (i) some or all may have been {\it accreted} from a high-mass infalling object; and (ii) we estimate an upper limit to contamination by genuine disk stars even at these high RVs at a level that does not allow this association to be conclusive. 
The remainder of the high-resolution stars have high [Ti/Fe] at low [Fe/H] and could be part of any of the {\it in-situ-halo}, {\it kicked-out} or the {\it accreted} populations.

\subsection{Our results in the context of other studies}
 
While it has had a long history \citep{hartwick87,zinn93,zinn96,sommerlarsen97}, the discussion of possible multiple mechanisms for the formation of the stellar halo has been revitalized in recent years both by the advent of large photometric data sets
with follow-up low-resolution spectroscopic work \citep[e.g.,][]{carollo07,carollo10}, and more modest, but ever-expanding samples of nearby halo stars with high-resolution spectroscopy \citep[e.g.,][]{nissen10,nissen11,schuster12,ishigaki12}.
In several cases, the studies point out classes of halo populations with distinct properties that are argued to match expectations for the properties of populations with distinct formation histories seen in hydrodynamical simulations of galaxy formation: the models generally predict an inner halo (within $\sim$20-30 kpc) dominated by metal-rich stars formed within the main Galactic progenitor and an outer halo dominated by lower-metallicity, accreted stars \citep{abadi06,zolotov09,mccarthy12}.
However, comparisons of the models to the data sets up to this point are necessarily inconclusive because the results of the simulations themselves are dependent on the (hard to model!) details of when, how, and where stars form in the main Galactic progenitor.

For example, \citet{carollo07,carollo10} claim evidence for two populations in their sample of 10,123 nearby (within 4 kpc) SDSS calibration stars: one of metal-rich stars (with a metallicity distribution peaking around [Fe/H]= -1.6) on only mildly eccentric orbits and a second of metal-poor stars on more eccentric orbits.
These observations are very reminiscent of the hydrodynamical simulation results --- indeed \citet{carollo10} {\it interpret} their metal-rich stars as an inner, {\it in-situ-halo}/{\it kicked-out} population and their metal-poor stars as an outer, {\it accreted} population --- but a transition in the orbital and metallicity properties in these populations is not necessarily inconsistent with a purely accreted stellar halo.
Moreover, the \citet{carollo07,carollo10} interpretation is at odds with our finding of a large fraction of clearly accreted stars in our M giant sample of even higher metallicity stars ([Fe/H] $>$ -1.2) than their metal-rich population.
In fact, we might expect the M giants to be biased {\it towards} finding {\it in-situ-halo}/{\it kicked-out} stars because late-type giants are a metal-rich stellar population.
On the other hand,
the selection of RV outliers admits a bias in our sample towards the high velocity tails of all populations --- 
particularly {\it accreted} stars, which are expected typically to be on higher energy and higher eccentricity orbits than the other stars.
This could explain why our sample is particularly sensitive to the accreted population.
Further work is needed to definitively show which of these two biases should dominate in a sample such as ours. 

Our results are more similar to those of \citet{nissen10,nissen11} and \citet{schuster12}, who find roughly equal-size populations distinct in age, abundances and orbits when they divide their (metal-rich, [Fe/H] $>$ -1.6) stellar halo sample between low-$\alpha$ and high-$\alpha$ stars in the [Mg/Fe]-[Fe/H] plane. 
They interpret the properties of the low-$\alpha$ population (in which the stars are found also to be younger and on more eccentric orbits) as being consistent with an accreted population --- so that these authors find an accreted fraction for the stellar halo at these metallicities similar to our own estimates. 
It is interesting to note that we also find stars with significantly lower [$\alpha$/Fe] (less than zero) than any of the stars in the \citet{nissen10} sample, possibly because our large survey volume (out to 9 kpc from the Sun) encompasses debris from more chemically extreme accreted progenitors not represented in their local sample (which is limited to within 335 pc of the Sun).

\subsection{Conclusion}
We explore the properties of relatively nearby, RV-selected halo M giant stars and conclude that the chemical properties of the stars in this sample show tentative evidence of distinct populations with distinct formation mechanisms. 
While close to 50\% of the stars fall in the {\it accreted} region of chemical abundance space, a definitive assessment of the relative contributions from {\it in-situ-halo}/{\it kicked-out} stars is not possible, due to the sometimes ambiguous categorizations of stars based upon their [$\alpha$/Fe]-[Fe/H] abundance distributions alone.
\citet{nissen11} have demonstrated that using abundance patterns along 
with more complete orbital information can play a key role in making this identification.
Our own chemodynamical data point to the importance of mapping a significant volume of the halo to confirm that such local studies are representative of global properties.
Follow-up work on the kinematics and more detailed chemical characterization of these stars would  give more insight into their origin.
Quantifying the size of the {\it kicked-out} population more generally could provide valuable constraints on the hydrodynamical models of galaxy formation.

A key finding is that our selection of M giants with unusual, halo-like RVs picks out a stellar halo population dominated by {\it accreted} stars.
Hence, searching for accretion events in RV-outlying samples of M giants should be prolific and motivates further interest in the putative groupings of stars found in our RV survey.
In a companion paper we explore what more the kinematical properties of these groupings could be telling us about their origins \citep{johnston12}.

\acknowledgements
The authors thank the referee for his/her helpful comments.  Many thanks to Eric Bell, Paul Harding, and Tim Beers for elucidating discussions at the January 2012 meeting of the American Astronomical Society. 
A.A.S. gratefully acknowledges support from The Vassar College Committee on Research and the Columbia Science Fellows Program. 
The work of K.V.J. and A.A.S. on this project was funded in part by NSF grants AST-0806558 and AST-1107373.
S.R.M. acknowledges partial funding of this work from NSF grant AST-0807945 and NASA/JPL contract 1228235.

\clearpage

\begin{deluxetable}{c c c c c c c c c c}
\rotate
\tabletypesize{\scriptsize}
\tablewidth{0pt}
\tablecaption{\label{tab2}Properties of the High-Resolution Program Stars.}
\tablehead{
\colhead{ID} & \colhead{$l$} & \colhead{$b$} & \colhead{($K_{S})_{0}$} & \colhead{$(J-K_{S})_{0}$} & \colhead{Date} & \colhead{Inst} & \colhead{S/N} & \colhead{$v_{\rm hel,h}/v_{\rm hel,m}$} & \colhead{origin} \\
 & \colhead{$deg$} & \colhead{$deg$} &  &  & \colhead{UT} &  &   & \colhead{km s$^{-1}$} &  \\
} 
\startdata
1521021+082320 &    12.1085 &   49.9949 &  7.458 &  0.969 & 01-Mar-2010  & ECHLR  &   62 &   -192.9/-183.3  & a \\
1546044+061554 &    14.3431 &   43.6135 &  7.745 &  0.961 & 26-Feb-2010  & ECHLR  &   75 &   -268.6/-270.6  & a \\
1535178+135331 &    22.2041 &   49.6287 &  6.569 &  1.028 & 30-Mar-2009  & ARCES  &  114 &     53.9/61.1 &    d \\
1640358+060251 &    22.6837 &   31.7562 &  5.655 &  1.013 & 30-Mar-2009  & ARCES  &   80 &     35.4/49.4 &    d \\
1509307+252912 &    37.8279 &   59.1097 &  8.680 &  0.991 & 25-Feb-2010  & ECHLR  &   84 &   -231.6/-231.6 &    a \\
1546468+270052 &    43.1496 &   51.1734 &  6.965 &  0.972 & 26-Feb-2010  & ECHLR  &   98 &   -217.3/-219.6 &    a \\
1530257+323409 &    51.7506 &   55.2995 &  5.997 &  1.053 & 30-Mar-2009  & ARCES  &  138 &     -8.6/-4.6 &   d \\
1719596+301146 &    53.2262 &   31.8831 &  6.863 &  0.919 & 02-Mar-2010  & ECHLR  &   52 &    -30.1/-27.0 &    d \\
1731343+401543 &    65.3999 &   31.7765 &  5.335 &  1.019 & 26-Feb-2010  & ECHLR  &  111 &    -28.2/-27.8 &    d \\
0913092+450916 &   175.2015 &   43.4028 &  8.045 &  0.937 & 26-Feb-2010  & ECHLR  &  105 &   -107.0/-108.2 &    a \\
0807206+435418 &   176.0975 &   31.6497 &  6.552 &  0.937 & 26-Feb-2010  & ECHLR  &  123 &   -122.5/-131.7 &    ko \\
0935556+414046 &   179.7566 &   47.7631 &  6.179 &  0.920 & 26-Feb-2010  & ECHLR  &  174 &   -252.7/-248.3 &    a \\
0903471+414944 &   179.7632 &   41.7691 &  6.322 &  0.955 & 26-Feb-2010  & ECHLR  &  195 &   -204.8/-210.6 &    a \\
0830244+283812 &   194.9298 &   33.0562 &  6.585 &  0.973 & 25-Feb-2010  & ECHLR  &  174 &    264.5/269.9 &    a \\
1011498+230504 &   210.2972 &   53.7906 &  7.908 &  0.934 & 02-Mar-2010  & ECHLR  &   89 &    238.4/242.4 &    a \\
0918573+145749 &   215.4937 &   39.3263 &  8.780 &  0.908 & 02-Mar-2010  & ECHLR  &   79 &    229.9/235.2 &    is/ko/a \\
0938204+112950 &   222.2501 &   42.1710 &  8.349 &  0.980 & 25-Feb-2010  & ECHLR  &   97 &    264.2/265.6 &    a \\
1001593+114507 &   225.6258 &   47.4509 &  8.399 &  0.932 & 02-Mar-2010  & ECHLR  &   77 &    179.2/183.2 &    is/ko/a \\
0932038+055521 &   227.7177 &   38.1643 &  7.397 &  1.073 & 02-Mar-2010  & ECHLR  &   82 &    190.3/171.7 &    ko \\
0950147+082356 &   227.8484 &   43.3082 &  4.981 &  1.154 & 02-Mar-2010  & ECHLR  &  107 &    143.8/167.4 &    ko \\
1101590+084043 &   243.2806 &   58.2289 &  8.748 &  0.931 & 02-Mar-2010  & ECHLR  &   73 &    263.8/263.9 &    a \\
1024571+004900 &   243.5677 &   46.1144 &  7.173 &  1.054 & 25-Feb-2010  & ECHLR  &  109 &    304.5/314.1 &    a \\
1101024-003004 &   254.5704 &   51.6925 &  7.208 &  0.923 & 25-Feb-2010  & ECHLR  &  172 &    264.9/267.0 &    is/ko/a \\
1115376+000800 &   258.5365 &   54.5283 &  8.248 &  1.037 & 02-Mar-2010  & ECHLR  &   72 &    196.4/194.6 &    ko \\
1054035-065124 &   258.7632 &   45.7090 &  7.355 &  1.024 & 25-Feb-2010  & ECHLR  &  154 &    267.1/269.9 &    is/ko/a \\
1037414-121048 &   259.0329 &   39.0328 &  5.977 &  1.142 & 25-Feb-2010  & ECHLR  &  185 &    248.2/257.1 &    ko \\
1136527+025949 &   263.2681 &   59.9961 &  7.387 &  0.988 & 25-Feb-2010  & ECHLR  &  117 &    162.1/159.3 &    a \\
1145072+013727 &   268.2122 &   59.9476 &  7.274 &  1.029 & 26-Feb-2010  & ECHLR  &  105 &    146.6/145.9 &    a \\
1105038-164000 &   269.3479 &   39.1698 &  7.386 &  0.980 & 02-Mar-2010  & ECHLR  &   89 &    349.7/354.6 &    is/ko/a \\
1131243-055825 &   269.6677 &   51.6539 &  6.526 &  1.158 & 26-Feb-2010  & ECHLR  &  129 &    123.2/137.9 &    is/ko/a \\
1206183-035045 &   281.8293 &   57.1651 &  6.521 &  1.095 & 25-Feb-2010  & ECHLR  &  151 &    220.6/238.9 &    a \\
1313176-164220 &   310.4547 &   45.8468 &  8.848 &  0.942 & 01-Apr-2009  & ARCES  &   82 &    113.7/96.9 &    ko \\
1314125-132352 &   311.4065 &   49.0990 &  7.125 &  0.943 & 01-Apr-2009  & ARCES  &  132 &     99.3/98.9 &    ko \\
1334567-055158 &   322.2190 &   55.3680 &  6.968 &  1.035 & 01-Apr-2009  & ARCES  &  100 &     71.5/77.6 &    ko \\
1355160-053350 &   330.5657 &   53.8480 &  7.352 &  1.057 & 02-Mar-2010  & ECHLR  &   86 &     91.9/96.2 &    is/ko/a \\
1442035-162350 &   337.8142 &   38.8681 &  7.298 &  1.050 & 02-Mar-2010  & ECHLR  &   73 &    152.2/145.9 &    is/ko/a \\
1404515-004157 &   338.2993 &   57.0434 &  6.590 &  1.109 & 02-Mar-2010  & ECHLR  &   73 &    114.9/121.4 &    is/ko/a \\
1431302-051730 &   343.4025 &   49.5528 &  8.156 &  0.909 & 01-Apr-2009  & ARCES  &   62 &    162.0/156.5 &    a \\
1518065-115536 &   350.0932 &   37.1732 &  7.588 &  1.045 & 30-Mar-2009  & ARCES  &   91 &    185.2/183.0 &    a \\
\enddata 
\end{deluxetable}

\begin{deluxetable}{ccccccccccccccc}
\tablecaption{\label{tab3}Atomic Lines and Equivalent Widths.}
\tabletypesize{\scriptsize}
\rotate
\tablehead{
\colhead{ID} & \colhead{7443.02} & \colhead{7447.38} & \colhead{7461.52} & \colhead{7498.53} & \colhead{7507.26} & \colhead{7511.02} & \colhead{7531.14} & \colhead{7540.43} & \colhead{7547.91} & \colhead{7568.89} & \colhead{7583.79} & \colhead{7474.94} & \colhead{7489.57} & \colhead{7496.12}\\
 & \colhead{\AA} & \colhead{\AA} & \colhead{\AA} & \colhead{\AA} & \colhead{\AA} & \colhead{\AA} & \colhead{\AA} & \colhead{\AA} & \colhead{\AA} & \colhead{\AA} & \colhead{\AA} & \colhead{\AA} & \colhead{\AA} & \colhead{\AA} \\
 & \colhead{FeI} & \colhead{FeI} & \colhead{FeI} & \colhead{FeI} & \colhead{FeI} & \colhead{FeI} & \colhead{FeI} & \colhead{FeI} & \colhead{FeI} & \colhead{FeI} & \colhead{FeI} & \colhead{TiI} & \colhead{TiI} & \colhead{TiI} \\
}
\tablewidth{0pt}
\startdata
1521021+082320 &    53.5 &    34.5 &    87.7 &     ... &     ... &   162.4 &   100.1 &    56.9 &    20.0 &    94.2 &   147.0 &    26.7 &    56.5 &    42.1 \\
1546044+061554 &    50.0 &    34.2 &    86.1 &    33.8 &    66.2 &   171.6 &    98.2 &    51.0 &     ... &    92.0 &   144.4 &    27.9 &    48.6 &    33.5 \\
1535178+135331 &    54.3 &    52.1 &   104.5 &    51.7 &     ... &   168.5 &   110.3 &    58.3 &    32.0 &    99.0 &   156.6 &    26.7 &   104.7 &     ... \\
1640358+060251 &    83.4 &    59.5 &   109.8 &    65.6 &   101.0 &   193.0 &   120.8 &    67.5 &    31.0 &   108.4 &   159.8 &    77.7 &   113.8 &    91.6 \\
1509307+252912 &    40.1 &    31.2 &    80.4 &    28.1 &    52.6 &     ... &    94.2 &    39.2 &    17.8 &     ... &   135.4 &    17.2 &    37.5 &     ... \\
1546468+270052 &    64.0 &    36.8 &    98.5 &    45.4 &    87.2 &   180.6 &   115.3 &    58.1 &    31.5 &   108.9 &   157.5 &    43.2 &    72.4 &    57.5 \\
1530257+323409 &    63.0 &    47.1 &   103.8 &    52.6 &    70.6 &   171.1 &   101.3 &    60.0 &    24.2 &    94.6 &   147.0 &    89.0 &   115.0 &   102.5 \\
1719596+301146 &    71.3 &    57.0 &   101.4 &     ... &    89.8 &   174.5 &   114.9 &    65.0 &    31.8 &   109.5 &   155.2 &    63.2 &   117.7 &    86.0 \\
1731343+401543 &    88.0 &    60.4 &   130.1 &    71.4 &   108.5 &   201.0 &   130.8 &    76.0 &    35.9 &   123.2 &   175.6 &    95.7 &   138.2 &   107.8 \\
0913092+450916 &    62.9 &    41.6 &    90.3 &    40.9 &    77.0 &   166.5 &   108.0 &    47.9 &     ... &    99.8 &   143.9 &    27.0 &    60.8 &    43.0 \\
0807206+435418 &     ... &    60.0 &   131.2 &    77.0 &   113.3 &   207.9 &   137.5 &    79.0 &    45.4 &   130.1 &   186.3 &    99.9 &   136.2 &   110.8 \\
0935556+414046 &    46.2 &    26.3 &    75.1 &    27.3 &    59.4 &   154.2 &    88.3 &    45.0 &    19.9 &    82.2 &   130.8 &    15.4 &    35.9 &    25.5 \\
0903471+414944 &    60.0 &    43.5 &    91.8 &    50.1 &    77.0 &   161.1 &   102.0 &    47.3 &    20.0 &    96.4 &   140.8 &    54.2 &    83.8 &    70.3 \\
0830244+283812 &    49.0 &    34.5 &    85.0 &    36.3 &    66.3 &   159.3 &    98.1 &    46.8 &    14.2 &    91.8 &   145.1 &    27.4 &    52.7 &    48.4 \\
1011498+230504 &    55.1 &    34.0 &    94.5 &    40.0 &    73.0 &   168.0 &   114.0 &     ... &    21.0 &    93.1 &   153.9 &    30.7 &    63.0 &    49.0 \\
0918573+145749 &    53.6 &    21.1 &    76.1 &     ... &    64.0 &   163.1 &    88.7 &    36.7 &    12.6 &    88.3 &     ... &    19.1 &    39.3 &    22.9 \\
0938204+112950 &    60.5 &    37.4 &    94.6 &    39.1 &    78.5 &   165.4 &   107.3 &    54.4 &    21.6 &   105.4 &     ... &    37.1 &    76.5 &    65.4 \\
1001593+114507 &    66.8 &    44.3 &    94.3 &     ... &    87.0 &   182.6 &   109.5 &    55.0 &    17.4 &    97.3 &   150.9 &    53.6 &    93.0 &    83.2 \\
0932038+055521 &    67.5 &    45.0 &    94.1 &    43.8 &    74.7 &   162.9 &   102.3 &    59.7 &    33.8 &    98.7 &   149.0 &    77.5 &   105.7 &    97.8 \\
0950147+082356 &    55.6 &    49.5 &   105.5 &    47.3 &     ... &   156.7 &     ... &    55.9 &    22.6 &    90.2 &   140.3 &    84.7 &   109.2 &    96.1 \\
1101590+084043 &    41.9 &     ... &    71.8 &     ... &    60.0 &   143.0 &   100.7 &    44.2 &    18.7 &    80.6 &   147.2 &     ... &    43.8 &    33.0 \\
1024571+004900 &    62.5 &    37.3 &   105.8 &    41.6 &    74.3 &   169.5 &   114.8 &    60.0 &    21.7 &    95.4 &   165.0 &    52.8 &    84.5 &    69.0 \\
1101024-003004 &    42.2 &    25.1 &    81.8 &    24.8 &    61.3 &   158.7 &    92.3 &     ... &     ... &    80.4 &   139.1 &    16.5 &    41.0 &    30.4 \\
1115376+000800 &    63.6 &    45.5 &    95.3 &     ... &    74.0 &   162.8 &     ... &    60.2 &    26.5 &    96.4 &   145.5 &    74.2 &   104.7 &    94.3 \\
1054035-065124 &    59.1 &    35.4 &    97.0 &    40.8 &    74.0 &   158.0 &    93.8 &    53.7 &    17.5 &    89.1 &   147.6 &    58.2 &    95.4 &    81.4 \\
1037414-121048 &    55.8 &    45.0 &   102.7 &    39.6 &    70.3 &   151.8 &    96.6 &    63.0 &    21.2 &    94.2 &     ... &    85.2 &   105.4 &    93.5 \\
1136527+025949 &    44.5 &    26.3 &    81.3 &    27.4 &    66.2 &   162.8 &    95.9 &    35.8 &    15.6 &    85.6 &   146.6 &    20.6 &    45.5 &    35.7 \\
1145072+013727 &    55.6 &     ... &    99.9 &     ... &    85.2 &   171.2 &   110.0 &    60.9 &    25.5 &    93.4 &   145.5 &    67.2 &    92.8 &    82.7 \\
1105038-164000 &    48.0 &    20.9 &    86.2 &    21.7 &    76.2 &   175.3 &   106.6 &    37.9 &     ... &    87.3 &   150.1 &    30.4 &    62.1 &    45.5 \\
1131243-055825 &    45.2 &    28.2 &    96.0 &    29.1 &    67.3 &   134.5 &     ... &    50.2 &    21.9 &    71.7 &   132.1 &    84.8 &   107.2 &    93.2 \\
1206183-035045 &    49.6 &    36.9 &    94.3 &    42.5 &    72.4 &   158.5 &   101.6 &    56.5 &    19.6 &    99.7 &   152.4 &    60.0 &    91.7 &    80.0 \\
1313176-164220 &    66.9 &    50.1 &   103.6 &    53.5 &    87.3 &   194.6 &   115.5 &    66.3 &    43.8 &   103.6 &   175.1 &    72.3 &   110.0 &    96.8 \\
1314125-132352 &    66.7 &    43.0 &    93.0 &    52.1 &    84.1 &   173.5 &   106.4 &    56.5 &    26.7 &    97.2 &   147.2 &    58.7 &    95.3 &    75.1 \\
1334567-055158 &    65.8 &    50.7 &    95.7 &    48.5 &    79.9 &   170.6 &   110.7 &    67.2 &    25.2 &    97.1 &   149.7 &    85.2 &   113.7 &    98.3 \\
1355160-053350 &    50.0 &    32.0 &    73.1 &    25.1 &    57.0 &   146.5 &    89.3 &     ... &     ... &    80.6 &   149.4 &    53.8 &    69.0 &    64.4 \\
1442035-162350 &    42.5 &    33.4 &    96.3 &    32.4 &    64.9 &   154.1 &    91.9 &    46.2 &    13.3 &    87.3 &   133.6 &    40.5 &    74.6 &    60.9 \\
1404515-004157 &    56.2 &    33.2 &   112.4 &    45.1 &    70.6 &   171.2 &   113.5 &    63.2 &    20.6 &   104.3 &   167.4 &    84.1 &   113.7 &    91.7 \\
1431302-051730 &    46.8 &    28.5 &    90.5 &    31.3 &    70.3 &   164.4 &   104.6 &    39.5 &     ... &     ... &   143.6 &     ... &    39.2 &    31.1 \\
1518065-115536 &    55.7 &    44.5 &    98.3 &    41.9 &    75.6 &   161.5 &   103.4 &    56.7 &     ... &   100.0 &   136.8 &    55.8 &    84.8 &    69.4 \\
\tableline
Arcturus (APO)  & 72.0 & 45.9 &  91.7 & ...  &  96.1 & 193.4 & 116.9 & 56.5 & 29.4 & 108.1 & 150.8 &  38.1 & 77.7   & 57.6 \\ 
Arcturus (KPNO) & 65.5 & 45.5 &  94.6 & 31.0 &  89.3 & 185.0 & 115.8 & 48.0 & 31.6 & 106.6 & 148.4 &  35.9 & 72.9   & 54.1 \\ 
$\alpha$ Tau (KPNO n1) & 90.3 & 63.4 & 120.8 & 62.3 & 115.4 & 212.7 & 132.4 & 77.3 & 45.3 & 129.0 & 179.0 & 99.6 & 127.9  & 109.1 \\ 
$\alpha$ Tau (KPNO n4) & 95.7 & 65.8 & 120.0 & 73.5 & 113.9 & 211.0 & 132.2 & 78.5 & 45.8 & 121.6 & 181.0 & 90.4 & 124.4 & 104.0 \\ 
$\delta$ Vir (APO) &  83.0 & 62.7 & 119.9 & 73.8 & 103.1 & 193.3 & ... & 76.4 & 40.1 & 112.2 & 164.8 & 99.3 & 126.8 & 121.3 \\
$\delta$ Vir (KPNO) & 76.1 & 54.3 & 117.4 & 68.0 & 95.6 & 188.5 &  ... & 69.7 & 44.4 & 112.0 & 164.2 & 99.9 & 131.5 & 120.6 \\
$\nu$ Vir (APO)  & 73.4 & 50.7 & 101.7 & 59.2 & 85.7 & 171.8 & 110.7 & 60.9 & 29.3 & 98.7  & 155.4 & 76.3 & 109.3 & 93.4 \\
$\nu$ Vir (KPNO) & 66.9 & 51.2 &  97.0 & 57.2 & 82.8 & 175.3 & 107.4 & 60.3 & 25.4 & 93.0  & 149.9 & 76.4  & 103.9 & 92.5 \\
\enddata
\end{deluxetable}

\begin{deluxetable}{ccccccccccccccc}
\tabletypesize{\scriptsize}
\rotate
\tablecaption{\label{tab4}Derived Properties of the Program Stars.}
\tablehead{
\colhead{ID} & \colhead{$T_{\rm eff}$} & \colhead{log $g$} & \colhead{$\xi$} & \colhead{$A$(Fe)} & \colhead{[Fe/H]} & \colhead{e[Fe/H]} & \colhead{$A$(Ti)} & \colhead{[Ti/Fe]} & \colhead{e[Ti/Fe]} & \colhead{dist} \\
 & \colhead{(K)} &  & \colhead{(km s$^{-1}$)} &  &  &  &  &  &  & \colhead{(kpc)} \\
}
\tablewidth{0pt}
\startdata
1521021+082320 &   3900 &   0.4 &  1.49 &   6.50 &  -0.95 &   0.07 &   3.85 &  -0.10 &   0.11 &   5.1 \\
1546044+061554 &   3900 &   0.5 &  1.58 &   6.44 &  -1.01 &   0.11 &   3.76 &  -0.13 &   0.17 &   5.9 \\
1535178+135331 &   3800 &   0.8 &  1.38 &   7.10 &  -0.35 &   0.14 &   4.14 &  -0.41 &   0.45 &   2.3 \\
1640358+060251 &   3800 &   0.9 &  1.54 &   7.30 &  -0.15 &   0.16 &   4.61 &  -0.14 &   0.04 &   1.3 \\
1509307+252912 &   3900 &   0.3 &  1.33 &   6.34 &  -1.11 &   0.13 &   3.56 &  -0.23 &   0.14 &   8.6 \\
1546468+270052 &   3900 &   0.6 &  1.62 &   6.77 &  -0.68 &   0.09 &   4.08 &  -0.14 &   0.14 &   3.3 \\
1530257+323409 &   3800 &   0.6 &  1.38 &   6.90 &  -0.55 &   0.15 &   4.70 &   0.35 &   0.12 &   2.2 \\
1719596+301146 &   3900 &   1.2 &  1.38 &   7.30 &  -0.15 &   0.10 &   4.74 &  -0.01 &   0.12 &   1.8 \\
1731343+401543 &   3800 &   1.1 &  1.72 &   7.44 &  -0.01 &   0.14 &   4.88 &  -0.01 &   0.04 &   0.9 \\
0913092+450916 &   3950 &   0.7 &  1.48 &   6.74 &  -0.71 &   0.10 &   3.96 &  -0.23 &   0.08 &   4.9 \\
0807206+435418 &   3900 &   1.3 &  1.72 &   7.52 &   0.07 &   0.13 &   5.01 &   0.04 &   0.07 &   1.3 \\
0935556+414046 &   4000 &   0.5 &  1.33 &   6.37 &  -1.08 &   0.11 &   3.73 &  -0.09 &   0.13 &   2.6 \\
0903471+414944 &   3900 &   0.6 &  1.42 &   6.71 &  -0.74 &   0.14 &   4.29 &   0.13 &   0.13 &   2.5 \\
0830244+283812 &   3950 &   0.2 &  1.55 &   6.37 &  -1.08 &   0.10 &   3.95 &   0.13 &   0.18 &   3.2 \\
1011498+230504 &   4000 &   0.7 &  1.63 &   6.57 &  -0.88 &   0.08 &   4.10 &   0.08 &   0.12 &   5.0 \\
0918573+145749 &   4000 &   0.4 &  1.79 &   6.19 &  -1.26 &   0.14 &   3.75 &   0.11 &   0.16 &   8.6 \\
0938204+112950 &   3900 &   0.5 &  1.56 &   6.63 &  -0.82 &   0.08 &   4.11 &   0.03 &   0.13 &   6.8 \\
1001593+114507 &   4000 &   0.5 &  1.75 &   6.50 &  -0.95 &   0.13 &   4.47 &   0.52 &   0.15 &   6.6 \\
0932038+055521 &   3700 &   0.9 &  1.20 &   7.33 &  -0.12 &   0.11 &   4.73 &  -0.05 &   0.13 &   1.6 \\
0950147+082356 &   3600 &   0.5 &  1.35 &   7.12 &  -0.33 &   0.16 &   4.52 &  -0.05 &   0.10 &   1.6 \\
1101590+084043 &   4000 &   0.4 &  1.47 &   6.31 &  -1.14 &   0.13 &   3.82 &   0.06 &   0.16 &   8.9 \\
1024571+004900 &   3800 &   0.3 &  1.65 &   6.57 &  -0.88 &   0.09 &   4.05 &   0.03 &   0.13 &   4.7 \\
1101024-003004 &   4000 &   0.3 &  1.59 &   6.21 &  -1.24 &   0.10 &   3.83 &   0.17 &   0.12 &   4.2 \\
1115376+000800 &   3800 &   0.6 &  1.32 &   6.91 &  -0.54 &   0.08 &   4.55 &   0.19 &   0.11 &   6.2 \\
1054035-065124 &   3800 &   0.0 &  1.42 &   6.46 &  -0.99 &   0.13 &   4.25 &   0.34 &   0.10 &   4.9 \\
1037414-121048 &   3650 &   0.1 &  1.28 &   6.82 &  -0.63 &   0.12 &   4.41 &   0.14 &   0.13 &   3.0 \\
1136527+025949 &   3950 &   0.4 &  1.69 &   6.27 &  -1.18 &   0.10 &   3.79 &   0.07 &   0.14 &   4.8 \\
1145072+013727 &   3800 &   0.4 &  1.49 &   6.71 &  -0.74 &   0.08 &   4.27 &   0.11 &   0.16 &   4.5 \\
1105038-164000 &   4000 &   0.3 &  2.05 &   6.12 &  -1.33 &   0.08 &   3.88 &   0.31 &   0.13 &   4.8 \\
1131243-055825 &   3650 &   0.0 &  1.19 &   6.54 &  -0.91 &   0.13 &   4.40 &   0.41 &   0.10 &   3.5 \\
1206183-035045 &   3700 &   0.1 &  1.47 &   6.57 &  -0.88 &   0.10 &   4.06 &   0.04 &   0.12 &   3.7 \\
1313176-164220 &   3900 &   1.0 &  1.58 &   7.13 &  -0.32 &   0.15 &   4.66 &   0.08 &   0.10 &   5.4 \\
1314125-132352 &   3950 &   0.9 &  1.40 &   6.92 &  -0.53 &   0.11 &   4.50 &   0.13 &   0.07 &   2.8 \\
1334567-055158 &   3750 &   0.8 &  1.36 &   7.16 &  -0.29 &   0.09 &   4.69 &   0.08 &   0.08 &   2.8 \\
1355160-053350 &   3800 &   0.1 &  1.47 &   6.28 &  -1.17 &   0.16 &   3.98 &   0.25 &   0.19 &   4.7 \\
1442035-162350 &   3800 &   0.2 &  1.50 &   6.36 &  -1.09 &   0.12 &   3.93 &   0.12 &   0.11 &   4.8 \\
1404515-004157 &   3700 &   0.1 &  1.76 &   6.56 &  -0.89 &   0.10 &   4.23 &   0.22 &   0.12 &   3.6 \\
1431302-051730 &   4000 &   0.5 &  1.69 &   6.36 &  -1.09 &   0.08 &   3.75 &  -0.06 &   0.19 &   6.8 \\
1518065-115536 &   3800 &   0.5 &  1.28 &   6.83 &  -0.62 &   0.10 &   4.19 &  -0.09 &   0.10 &   4.8 \\
\tableline
Arcturus (APO) 	 &	4250 &	1.4 &	1.73 &	6.88 &	-0.57 &	0.09 &	4.62 &	0.29 &	0.10 &  \\	
Arcturus (KPNO)	 &	4250 &	1.4 &	1.76 &	6.82 &	-0.63 &	0.09 &	4.56 &	0.29 &	0.11 &  \\	
$\alpha$ Tau(KPNO n1) &	3900 &	1.3 &	1.68 &	7.50 &	0.05 &	0.11 &	4.98 &	0.03 &	0.12 &  \\	
$\alpha$ Tau(KPNO n4) &	3900 &	1.3 &	1.50 &	7.62 &	0.17 &	0.13 &	4.99 &	-0.08 &	0.06 &  \\	
$\delta$ Vir(KPNO) &	3650 &	0.8 &	1.29 &	7.63 &	0.18 &	0.14 &	5.11 &	0.07 &	0.11 &  \\	
$\delta$ Vir(APO) &	3700 &	0.8 &	1.35 &	7.62 &	0.17 &	0.14 &	5.05 &	0.03 &	0.15 &  \\
$\nu$ Vir(KPNO) &	3800 &	0.8 &	1.40 &	7.11 &	-0.34 &	0.15 &	4.58 &	0.02 &	0.12 &  \\	
$\nu$ Vir(APO) &	3800 &	0.8 &	1.36 &	7.19 &	-0.26 &	0.14 &	4.64 &	0.03 &	0.07 &  \\	
\enddata
\end{deluxetable}

\begin{deluxetable}{ccccc}
\tabletypesize{\scriptsize}
\tablecaption{\label{SLcomp}Standard Star Chemical Abundance Comparisons}
\tablehead{
\colhead{Star} & \colhead{[Fe/H]} & \colhead{[Fe/H]$_{\rm S\&L}$} & \colhead{[Ti/Fe]} & \colhead{[Ti/Fe]$_{\rm S\&L}$} \\
}
\tablewidth{0pt}
\startdata
Arcturus & -0.60$\pm$0.09 & -0.67$\pm$0.11  & 0.29$\pm$0.11 & 0.41$\pm$0.12 \\
$\alpha$ Tau & 0.11$\pm$0.10 & 0.07  & -0.03$\pm$0.08 & -0.03 \\
$\nu$ Vir & -0.30$\pm$0.15 & -0.02$\pm$0.20 &  0.03$\pm$0.20 & 0.06$\pm$0.17 \\
$\delta$ Vir & 0.18$\pm$0.17 & 0.13$\pm$0.17 & 0.05$\pm$0.13 & 0.07$\pm$0.20 \\
\enddata
\end{deluxetable}

\end{document}